\documentclass{aa}
\usepackage{txfonts}
\usepackage{graphicx}
\usepackage{natbib}
\usepackage{amstext,cases}
\bibpunct{(}{)}{;}{a}{}{,}
\usepackage[dvips]{xcolor}

\begin{document}

\title{Crucial aspects of the initial mass function (I)}
\subtitle{The statistical correlation between the total mass of an ensemble of stars and its most 
massive star}

\author{Miguel Cervi{\~n}o$^{1,2}$,  Carlos Rom{\'a}n-Z{\'u}{\~n}iga$^{3}$, Valentina Luridiana$^{2,4}$, 
Amelia Bayo$^{5,6}$, N{\'e}stor S{\'a}nchez$^7$ and Enrique P{\'e}rez$^1$}
\institute{Instituto de Astrof{\'\i}sica de Andaluc{\'\i}a (IAA-CSIC), Glorieta de la Astronom{\'\i}a s/n, 18008 Granada, Spain
\and Instituto de Astrof{\'{\i}}sica de Canarias, c/ v{\'{\i}}a L\'actea s/n, 38205 La Laguna, Tenerife, Spain 
\and Instituto de Astronom{\'{\i}}a, Universidad Acad\'emica en Ensenada, Universidad Nacional Aut\'onoma de M\'exico,
Ensenada BC, 22860 Mexico
\and Departamento de Astrof{\'{\i}}sica, Universidad de La Laguna (ULL), 38205 La Laguna, Tenerife, Spain
\and European Southern Observatory, Casilla 19001, Santiago 19, Chile
\and Max Planck Institut f\"ur Astronomie, K\"onigstuhl 17, 69117, Heidelberg, Germany
\and S. D. Astronom\'{\i}a y Geodesia,
Fac. CC. Matem\'aticas,
Universidad Complutense de Madrid,
28040, Madrid,
Spain.
}
\offprints{M. Cervi\~no \email{mcs@iaa.es}}
\date{Received ; accepted }

\abstract
{Our understanding of stellar systems depends on the adopted interpretation of the initial mass function, IMF $\phi(m)$. Unfortunately, there is not a common interpretation of the IMF, which leads to different methodologies and diverging analysis of observational data.}
{We study the correlation between the most massive star that a cluster would host, $m_\mathrm{max}$, and its total mass into stars,  ${\cal M}$,  as an example where different views of the IMF lead to different results.}
{We assume that the IMF is a probability distribution function and analyze the $m_\mathrm{max}-{\cal M}$ correlation within this context. We also examine the meaning of the equation used to derive a theoretical ${\cal M}-\hat{m}_\mathrm{max}$ relationship, ${\cal N} \times \int_{\hat{m}_\mathrm{max}}^{\mathrm{m_{up}}} \phi(m)\,\mathrm{d}m = 1$ with $\cal N$ the total number of stars in the system, according to different interpretations of the IMF.}
{We find that only a probabilistic interpretation of the IMF,
where stellar masses are identically independent distributed random variables, 
provides a self-consistent result. Neither $\cal M$ nor the total number of stars in the cluster, $\cal N$, can be used as IMF scaling factors. In addition, $\hat{m}_\mathrm{max}$ is a characteristic maximum stellar mass in the cluster, but not the actual maximum stellar mass. 
A $\left<{\cal M}\right>-\hat{m}_\mathrm{max}$ correlation is a natural result of a probabilistic interpretation of the IMF; however, the distribution of observational data in the ${\cal N} ~Ê(\mathrm{or}~Ê{\cal M}) - m_\mathrm{max}$ plane includes a dependence on the distribution of the total number of stars, $\cal N$ (and $\cal M$), in the system, $\Phi_{\cal N}({\cal N})$, which is not usually taken into consideration.}
{We conclude that a random sampling IMF is not in contradiction to  a possible $m_\mathrm{max} - {\cal M}$ physical law. However, such a law cannot be obtained from IMF algebraic manipulation or included analytically in the IMF functional form. The possible physical information that would be obtained from the ${\cal N} ~Ê(\mathrm{or}~Ê{\cal M}) -m_\mathrm{max}$ correlation is closely linked with the $\Phi_{\cal M}({\cal M})$ and  $\Phi_{\cal N}({\cal N})$ distributions; hence it depends on the star formation process and the assumed definition of stellar cluster.}

\authorrunning{Cervi\~no et al.}
\titlerunning{The IMF  and the $m_\mathrm{max}-{\cal M}$ statistical correlation}
\maketitle

\section{Introduction}

In recent literature, the term initial mass function (IMF) is
used to indicate three different types of distributions:
(1) the distribution by number of the stellar masses observed in a 
particular star ensemble, (2) a normalized version of (1), i.e., the 
frequency distribution of the stellar masses observed in a 
particular star ensemble, and (3) the theoretical probability density 
function $\phi(m)$ of the stellar masses that can be formed in a generic 
star ensemble. In this work, following \cite{Sca86}, we adopt the third 
definition and explore some consequences of mixing these definitions.

In the following, we leave distribution (2) out of the discussion
and focus, for simplicity, only on distributions (1) and (3)\footnote{However, because 
distribution (2) is an scaled version of distribution (1), the conclusions derived from (1) also apply to (2).}.
These two distributions are different but closely related to each other, 
as statistics and probability are. Probability deals with predicting the likelihood of
possible events in a system with known properties; statistics consists in
analysing the distribution of real events with the aim 
of determining some unknown property of the system. Probability addresses 
the direct problem, while statistics addresses the inverse problem.
In our case, distribution (3) describes the underlying probability
distribution from which stellar masses can be drawn,
while distribution (1) describes an actual stellar sample from 
which we wish, ideally, to recover the parameters of the underlying probability
distribution.

The relation between the shape of (1) and the shape of (3) depends
crucially on the size of the sample, that is, the number of stars
${\cal N}$; when ${\cal N}$ values are large, the two shapes tend to be similar.
This similarity can mislead one into believing that (1) is just a scaled-up version of (3), 
with ${\cal N}$ being the scale factor. This would be very wrong since, as explained above, the physical meanings of both distributions are intrinsically different. This paper is dedicated to exploring the implications of such difference.

A major drawback of the distribution-by-number view (number (1) above) is that the very definition of a stellar sample
necessarily implies some (hidden or explicit) assumption on the star formation (SF) process that originated the sample.
For example, an embedded, open, or globular cluster, an OB associations, and so on, are coeval and cospatial samples;
field stars, which are used to study galaxy structure, are neither coeval nor cospatial;
the stars in a galaxy that were born at a given time, which are a sample suitable for stellar populations studies, 
are coeval but not cospatial.
These examples make clear that, when a sample is selected, some predefined spatial and time scales are 
implicitly assumed, and these scales may influence the distribution by the number of the stellar masses.
Rephrasing \cite{Sca86}, when talking about the IMF, we are left in the uncomfortable position of having no 
means to define an empirical sample that 
corresponds to a consistent definition of IMF and that can be directly related to the theories of 
SF without introducing major assumptions.

The probability distribution function (pdf) view (number (3) above) is actually an abstraction used to describe the general  
universe of initial masses that a star would have. This interpretation implies that we have to use a probability framework 
in order to make a description of the problem and inferences from observed data sets. One implicit requirement of 
such an approach is that the stellar mass is an identically independent distributed (iid) variable, and therefore, any realization of the 
IMF is a random sample\footnote{Random sample means that every possible sample has a calculable chance 
of selection. This is a requirement of any statistical 
and probabilistic study \cite[][]{KS04}.}. Within this framework, all the empirical samples are included naturally as far 
as they are particular realizations 
of the theoretical distribution. Although it is possible to include conditions representing particular SF scenarios, it is generally assumed 
that the IMF has no memory of the SF event: that is, the SF details have no major impact on the IMF itself, although 
they can have an impact on the resulting IMF 
realization once the corresponding conditions are included in the derivation. It is a surprising fact that there is no clear 
observational evidence that the IMF varies strongly and 
systematically as a function of different SF scenarios \cite[][]{BCM10}.

Throughout this paper, we consider several pieces of work based on a distribution-by-number 
interpretation of the IMF. The specific way in which the IMF is represented varies depending on the 
considered paper. Some authors assume that the IMF is a continuous law that returns, for each mass 
value, the number of stars of that mass; others consider that it returns the number of stars in each mass bin. 
Some assume that the stars are distributed in a predefined way and the mass of a star depends on the mass 
of the other stars; others consider that the stars are distributed independently from each other. In the following, 
we give examples of this and emphasize the differences between the various distribution-by-number 
interpretations and the pdf view of the IMF.

Naturally, the equations involving the IMF depend on the interpretation of the IMF. More importantly however, the cluster-related quantities
inferred from manipulations of the IMF are interpreted differently according to the initial assumptions. 
One case in which the different views of the IMF lead to dramatically diverging interpretations is the 
modeling of the correlation between the total stellar mass in a cluster, $\cal M$, and the mass, 
$m_\mathrm{max}$, of its most massive star, which we investigate in this series of papers.

There are many facets to the study of the ${\cal M}-m_\mathrm{max}$ correlation. One is the correlation obtained 
theoretically from manipulations 
of the IMF functional form, which is the subject of this paper. 
Another is the inference of ${\cal M}$ from partial information of the system. The lack of information makes 
this inference deeply dependent on the IMF interpretation \cite[this aspect is discussed in ][,~hereafter Paper II]{Ceretal12}. 
A third issue is the comparison between theory and observational data. This point also depends on the 
interpretation of the IMF (and is studied in Jimenez-Donaire et al 2013 in prep., from now on Paper III)

The structure of the paper is as follows: In Sect.~\ref{sec:formal} we present
our basic framework for a probabilistic interpretation of the IMF. Section~\ref{sec:Ntot-mmax} is
devoted to analyzing {\it in a probabilistic context} the meaning of the basic equation commonly used in the literature relating
${\cal M}$ and $m_\mathrm{max}$. 
In Sect.~\ref{sec:discusion} we discuss the different methodologies and assumptions 
used by other authors to obtain a ${\cal M}-m_\mathrm{max}$ correlation.
We include a discussion on iid stellar masses and on the connection of the IMF with the SF.
Finally, we briefly discuss the composition 
of different IMFs to obtain an integrated galaxy IMF (IGIMF). Our conclusions are described in Sect.~\ref{sec:conclusions}.

\section{Formal probabilistic formulation}
\label{sec:formal}

Let us start by framing the problem in a formal probabilistic framework: 

\begin{enumerate}
\item The IMF, $\phi(m)=\mathrm{d}N/\mathrm{d}m$, is a pdf, that provides  the 
probability of finding a star in a given mass range by its integration in such mass range. 
The mass limits of the pdf, $\mathrm{m_{low}}$ and $\mathrm{m_{up}}$, are given by
stellar theory and must fulfill $\int_{\mathrm{m_{low}}}^{\mathrm{m_{up}}} \phi(m) \mathrm{d}m = 1$; 
that is, we are certain that any possible star has a mass between $\mathrm{m_{low}}$ and $\mathrm{m_{up}}$. 
This is the first fundamental difference with respect to the distribution-by-number interpretation: the IMF 
cannot be arbitrarily normalized to $\cal M$ or  $\cal N$, since it does not provide numbers of stars with 
a given mass but the probability for a star to be born with a given mass {\it independently} of how many 
stars are in the cluster or the cluster total mass. In this interpretation of the IMF, there is neither an implicit 
sample nor predefined space or time scales.

The IMF so defined may have values larger than one, provided its integral over any mass range is lower than one. 
This is the second fundamental difference with respect to the distribution-by-number interpretation when described 
in terms of frequencies (case 2 in the Introduction) where no value larger than one is possible by construction.

In this paper 
we use the Kroupa IMF \citep{Kro01,Kro02} as 
used  in \cite{WK06}\footnote{We note that \cite{WK04} use
  $\alpha_2=2.30$ in their parametrization of the IMF and that \cite{WK06}
  use $\alpha_2=2.35$.} and subsequent works, except for the value of 
$\mathrm{m_{up}}$ which we set equal to
$120\mathrm{M}_\odot$. Although a larger value would probably be more realistic according to recent 
studies (\citealt{Croetal10}, see also the contributions to the {\it Up2010} conference published by \citealt{Teretal11}), 
this choice is motivated by the fact that the
$\mathrm{m_{up}}$ value of most public stellar tracks used in most $m_\mathrm{max}$ estimations 
is $120\mathrm{M}_\odot$.  
In Fig.~\ref{fig:imf} we show the $\phi(m)$ used in this paper and   the probability for a star of having a mass 
in the range $m,m+1 \mathrm{M}_\odot$.

The probability for a random star of having a mass {\it lower than} a given value $m_\mathrm{a}$ is given by

\begin{equation}
p(m < m_\mathrm{a}) =  \int_{\mathrm{m_{low}}}^{m_\mathrm{a}} \phi(m) \, \mathrm{d}m,
\label{eq:pltm}
\end{equation}

while the probability  for a random star of having a mass {\it equal to or larger than} 
$m_\mathrm{a}$ is given by

\begin{equation}
p(m \geq m_\mathrm{a}) =  \int_{m_\mathrm{a}}^{\mathrm{m_{up}}} \phi(m) \, \mathrm{d}m.
\label{eq:pgeqm}
\end{equation}

In this work, the integrals over the IMF will always be
read as {\it equal to or larger than} the lower limit and  {\it lower than} the upper limit.
The use of {\it lower than} instead of {\it equal to or lower than} in the upper limit and the 
complementary in the lower limit is just a convention. However, 
{\it equal} cannot be used simultaneously in both equations:  no star can
simultaneously belong to two independent intervals.
The convention we use implies that the nominal 
value $\mathrm{m_{up}}$ cannot be formally reached, although values very close to it are possible.

\begin{figure}
\resizebox{\hsize}{!}{\includegraphics{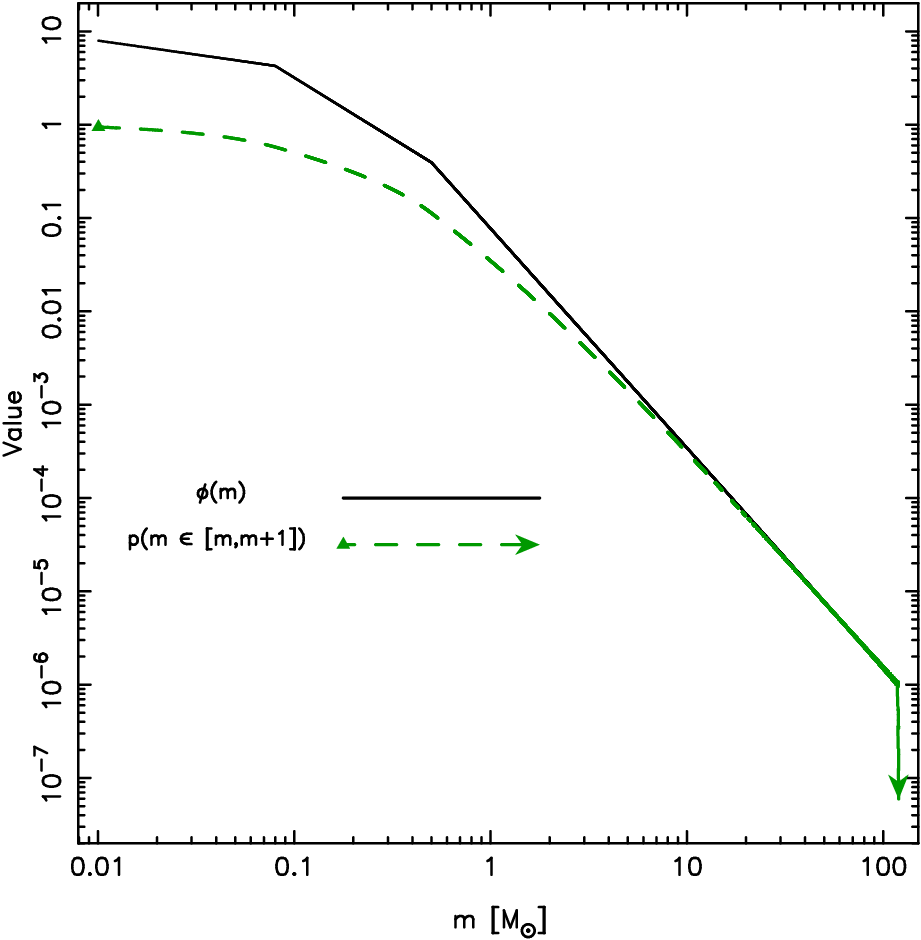}}
\caption[]{IMF used in the present work (solid line), as in the
  parametrization by \cite{Kro01,Kro02} and \cite{WK06}. 
  Being a pdf, it can have values larger than one; the probabilities
  are given by the integral over the pdf. We also plot the probability that a star has a 
  mass in the $m,m+1 \mathrm{M}_\odot$ range, which is lower than one (dashed line). 
  This probability declines rapidly when $m$ is larger than $\mathrm{m_{up}}-1 \mathrm{M}_\odot$. 
 }
\label{fig:imf}
\end{figure}

\item Different observational scenarios can be described by adding constraints to the IMF.  For instance, 
we may explicitly include the limit imposed on $m\mathrm{_{max}}$ by the total mass of the sample we 
are analyzing, that is,   $m\mathrm{_{max}} = \mathrm{min}\{\mathrm{m_{up}},
{\cal M}\}$. In this case, we must define an a posteriori pdf, related to the IMF, that includes such a condition:

\begin{eqnarray}
\phi(m | m < m\mathrm{_{max}}) = 
\frac{\phi(m) \, \mathrm{H}(m\mathrm{_{max}}-m)}{p(m <  m\mathrm{_{max}})},
\label{eq:IMFmcond}
\end{eqnarray}

\noindent where $\mathrm{H}(m_\mathrm{max} - m)$ is the Heaviside function\footnote{We use here the Heaviside 
function as a distribution to define the domain of $\phi(m)$, including constraints. In this situation the value of 
$\mathrm{H}(0)$ is {\it not} defined, but it is assigned  a posteriori to be consistent with the convention used 
in the integral limits. In the case of Eq.~\ref{eq:IMFmcond}, $\mathrm{H}(0)=0$.}, which ensures that no star 
equal to or larger than  $m_\mathrm{max}$ can be present in the cluster. 
We note that $\phi(m | m < m\mathrm{_{max}}) $ is also a pdf. The mean mass of such distribution is

\begin{equation}
 \left< m  | m < m\mathrm{_{max}} \right> = \frac{ \int_{\mathrm{m_{low}}}^{\mathrm{m_{up}}} \, 
 m \, \phi(m) \, \mathrm{H}(m\mathrm{_{max}}-m) \mathrm{d}m}{p(m < m\mathrm{_{max}})}.
 \label{eq:mmeancond}
\end{equation}

More elaborated constrained-IMF can be formulated, always keeping in mind that conditions are imposed ad hoc 
and produce a pdf whose functional form differs from $\phi(m)$.

\item The pdf describing ensembles with a total number of stars $\cal N$ 
(formally conditioned to have $\cal N$ stars) can be calculated as successive convolutions 
of the corresponding pdf for one star. 
For instance, the pdf  for the total mass, $\Phi_{\cal M}({\cal M}|{\cal N})$,  is 
the result of convolving the IMF ${\cal N}$ times with itself
\cite[see][]{CLCL06,SM08}:

\begin{equation}
\Phi_{{\cal M}}({\cal M}|{\cal N}) = \overbrace{ \phi(m)\otimes \phi(m) \otimes \, .... \,\otimes 
\phi(m)}^{{\cal N}}.
\label{eq:Mtot}
\end{equation}

A property of self-convolution is that  simple relations link  the mean value and the high-order moments 
of $\phi(m)$ and $\Phi_{{\cal M}}({\cal M}|{\cal N})$  \cite[see, e.g.,][]{CLCL06}. As an example, the mean 
integrated mass of $\Phi_{{\cal M}}({\cal M}|{\cal N})$, $ \left< {\cal M} | {\cal N} \right> $, is related to the 
mean stellar mass of the IMF, $ \left< m \right>$, through the relation

\begin{equation}
 \left< {\cal M} | {\cal N} \right> = {\cal N} \times \left< m \right> = {\cal N} \times 
 \int_{\mathrm{m_{low}}}^{\mathrm{m_{up}}} \, m \, \phi(m) \, \mathrm{d}m.
 \label{eq:meanvaluegen}
 \end{equation}
 
However, we note that $\Phi_{{\cal M}}({\cal M}|{\cal N}) \neq {\cal N}Ê\times \phi(m)$ and that the {\it actual} 
total mass cannot be obtained, but only an estimate of it.  This is the third fundamental difference with the 
distribution-by-number interpretation, which assumes that for a given ${\cal N}$ there is one, and only one, 
$\cal M$ value, given by  ${\cal M}({\cal N}) = {\cal N} \times \left< m \right>$.

\end{enumerate}

\section{Relating the number of stars with the most massive star in the sample}
\label{sec:Ntot-mmax}

According to the law of large numbers, in a sample of $\cal N$ stars drawn from an underlying pdf, $\phi(m)$, 
the typical number of stars $N_\mathrm{a}$  with $m \ge m_\mathrm{a}$ is given by 
$N_\mathrm{a} = {\cal N} \times p(m \geq m_\mathrm{a})$. Particularizing this equation,
we can {\it define} a characteristic maximum value of $m_\mathrm{max}$,  $\hat{m}_\mathrm{max}$, 
for which there is typically only one star with mass equal to or larger than  $\hat{m}_\mathrm{max}$ through

\begin{equation}
 1 = {\cal N} \times p(m \geq \hat{m}_\mathrm{max}) =  {\cal N} \times \int_{\hat{m}_\mathrm{max}}^\mathrm{m_{up}} \phi(m)\, \mathrm{d}m.
\label{eq:charval} 
\end{equation}

This is the basic equation used by several authors as the {\it determination} of the actual mass of the most 
massive star in a system \cite[as examples:][]{Elm97,Elm99,Elm00,KW03,WK04,WK06}. However, we can 
also obtain a mean value of $m_\mathrm{max}$ \citep{OC05} or a median value of $m_\mathrm{max}$ \citep{WKB10}. 
So the question is: does the definition of the characteristic value $\hat{m}_\mathrm{max}$ indeed provide the actual 
$m_\mathrm{max}$ extreme value or only an estimate of it? And if it is an estimate, what is its exact meaning? 
Let us seek the answer in a probabilistic context\footnote{The discussion in this section is mainly based on 
\cite{sornette}, \cite{KS04}, and  \cite{gumbel}, although the same formulae can be found in other works.}.

\begin{figure}
\resizebox{\hsize}{!}{\includegraphics{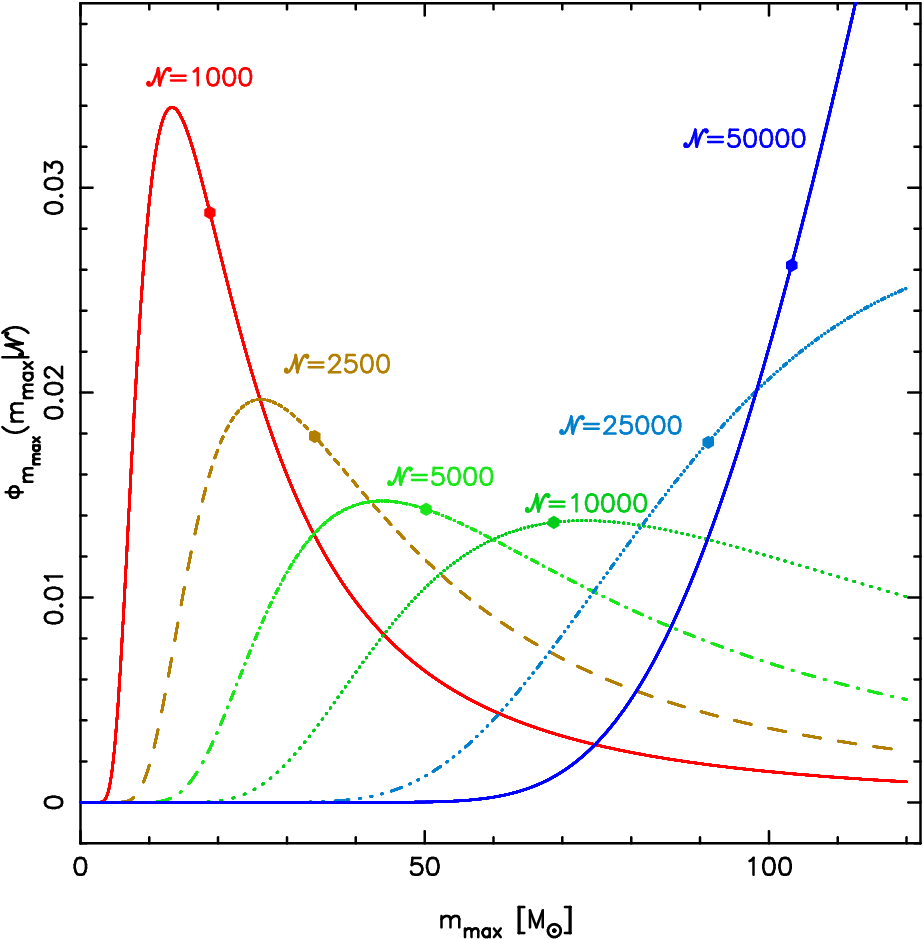}}
\caption[]{Distribution of the maximum stellar mass,  $\Phi_{m_\mathrm{max}}({m_\mathrm{max}}|{\cal N})$ 
for different values of ${\cal N}$.  The circle on each curve is the position of the characteristic value $\hat{m}_\mathrm{max}$.}
\label{fig:pdfmmax}
\end{figure}

We consider a set of ${\cal N}$ stars with unknown stellar masses, $m_i$, drawn from the IMF.  For any given 
mass $m_\mathrm{a}$, the probability of having {\it at least} one star with mass $m_i$ equal to or larger than $m_\mathrm{a}$ in the 
sample,  ${\cal{P}}(\exists i \in [1,{\cal N}]\, |\, m_i \geq m_\mathrm{a}) $, is the complementary probability  that all stars have a mass 
lower than $m_\mathrm{a}$, ${\cal{P}}(m_i < m_\mathrm{a}, \forall i \in \left[1,{\cal N}\right]) $. 
Since the stellar masses are iid drawn from the same 
distribution $\phi(m)$, the probability ${\cal{P}}(m_i < m_\mathrm{a}, \forall i \in \left[1,{\cal N}\right])$ is the result of multiplying 
$p(m < m_\mathrm{a})$ by itself  ${\cal N}$ times\footnote{Here we use $p$ to represent probabilities on the IMF 
(cf., Eqs.~\ref{eq:pltm} and \ref{eq:pgeqm}) and ${\cal{P}}$ to represent probabilities on the sample with ${\cal N}$ stars.}:

\begin{eqnarray}
{\cal{P}}(m_i < m_\mathrm{a}, \forall i \in \left[1,{\cal N}\right]) &=&  \left[p(m < m_\mathrm{a}) \right]^{{\cal N}} = \nonumber \\
&=& \left[1 - p(m \geq m_\mathrm{a}) \right]^{{\cal N}}.
\label{eq:P.lt.mmax}
\end{eqnarray}

\noindent Thus, 

\begin{eqnarray}
{\cal{P}}(\exists i \in [1,{\cal N}] |\, m_i \geq m_\mathrm{a})  &=& 1 - {\cal{P}}(m_j < m_\mathrm{a}, \forall j \in \left[1,{\cal N}\right]) = \nonumber\\
&=& 1-  \left[1 - p(m \geq m_\mathrm{a}) \right]^{{\cal N}}.
\label{eq:P.ge.mmax}
\end{eqnarray}

\noindent This relation is valid for any value of $m_\mathrm{a}$ and any distribution function. 

If we now set $m_\mathrm{a} = \hat{m}_\mathrm{max}$, we can replace $p(m \geq \hat{m}_\mathrm{max})$ in 
Eq.~\ref{eq:P.ge.mmax} by $1/{\cal N}$ by virtue of the $\hat{m}_\mathrm{max}$ definition. The probability that there is 
{\it at least} one star with $m\ge \hat{m}_\mathrm{max}$ in a sample of $\cal N$ stars is thus given by

\begin{equation}
{\cal{P}}(\exists i \in [1,{\cal N}] |\, m_i \geq \hat{m}_\mathrm{max})  =  1 - \left[1 - \frac{1}{{\cal N}} \right]^{{\cal N}},
\end{equation}

\noindent which has an asymptotic value $1 - 1/\mathrm{e} \sim 0.63$ for large ${\cal N}$ values, with 0.63 being 
a reasonable approximation for, say, ${\cal N} > 100$. Hence, the characteristic mass, $\hat{m}_\mathrm{max}$, obtained by solving 
Eq.~\ref{eq:charval} is the value of $m$ that is not reached or exceeded\footnote{We note that, depending on the reference and the 
convention used in Sect.~\ref{sec:formal},  this value can be defined either as reached or exceeded or just as exceeded.} with a 
probability 0.37 in a sample of $\cal N$ stars.
This means that in a large enough set of clusters, all of them with ${\cal N}$ stars, typically in 63\% of the clusters 
 the mass of the most massive star will be equal to {\it or larger} than $\hat{m}_\mathrm{max}$, while in 37\% of the clusters 
 it will be lower than $\hat{m}_\mathrm{max}$. So the $\hat{m}_\mathrm{max}$ value obtained in Eq.~\ref{eq:charval} does not provide 
 the mass $m\mathrm{_{max}}$ of the most massive star  in a cluster of ${\cal N}$ stars, contrary to what is stated in several 
 astrophysical papers\footnote{The characteristic largest value defined by Eq.~\ref{eq:charval} is related to
the estimation of the number of events we must record to have an event larger than a given value $m_\mathrm{a}$ (which is called {\it 
return period}  in extreme value theory). If the events are taken in a regular time interval, for instance, it could be the estimation of the 
number of years between earthquakes larger than a given magnitude, the number of years between economy crashes, and so on.}.
 
Actually, for any possible value $\hat{m}_\mathrm{max}$ lower than $\mathrm{m_{up}}$ that we would use as a proxy of the actual value of $m_\mathrm{max}$, there is a probability larger than 90\% that the most massive star in the system is more massive than such $\hat{m}_\mathrm{max}$ value (see Appendix \ref{subsubsect:IntFunc} for details).

\subsection{The pdf of $m_\mathrm{max}$ for a known $\cal N$, $\Phi_{m_\mathrm{max}}(m_\mathrm{max}|{\cal N})$ }
\label{subsec:Nfix}

\begin{figure}
\resizebox{\hsize}{!}{\includegraphics{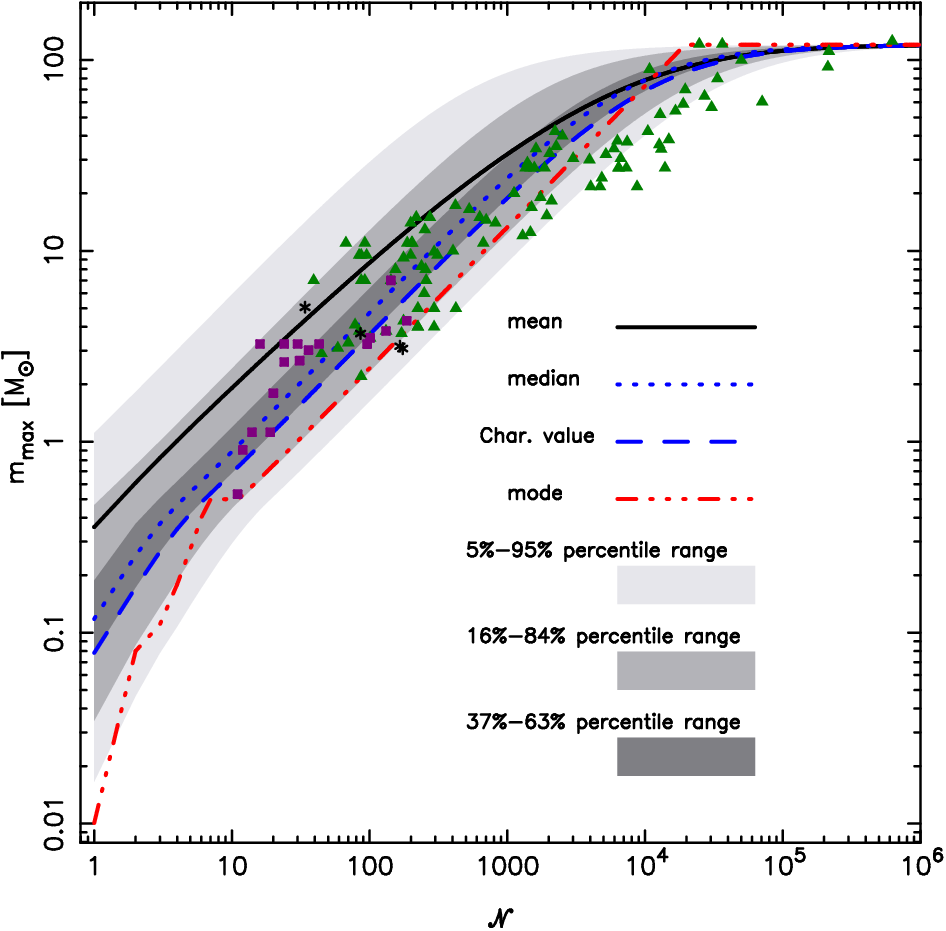}}
\caption[]{Percentile analysis around the median of  $\Phi_{m_\mathrm{max}}({m_\mathrm{max}}|{\cal N})$ as a function of ${\cal N}$ (shaded areas).  The figure includes as a reference the position of the characteristic value, median, mean, and mode of the distribution. Small triangles: compilation by \cite{WKB10} of observational values of $m_\mathrm{max}$ and inferred values of $\cal N$  obtained from observations; squares: observed values of $\cal N$ and $m_\mathrm{max}$ from \cite{KM11}; stars: observed values of $\cal N$ and $m_\mathrm{max}$ in the field for the four observed regions from \cite{KM11}.}
\label{fig:percentmmax}
\end{figure} 

\begin{figure}
\resizebox{\hsize}{!}{\includegraphics{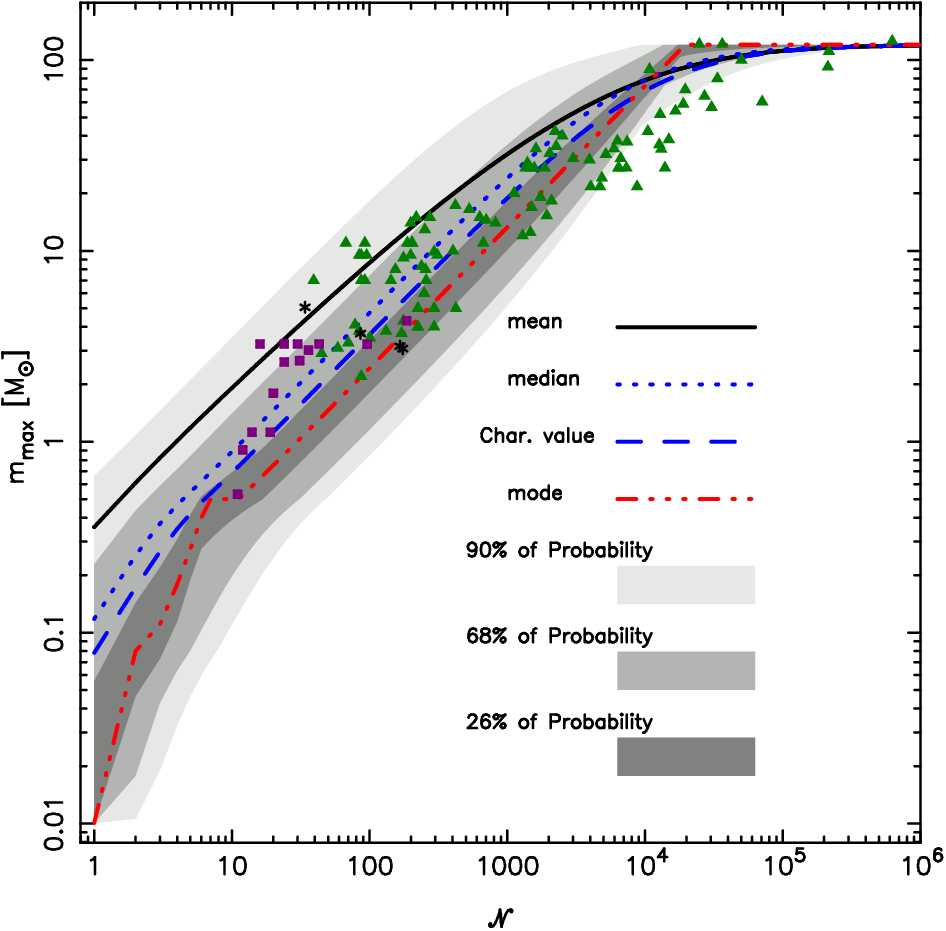}}
\caption[]{Confidence interval analysis of  $\Phi_{m_\mathrm{max}}({m_\mathrm{max}}|{\cal N})$ as a function of ${\cal N}$ (shaded area). Lines and symbols have the same meaning as in Fig.~\ref{fig:percentmmax}.}
\label{fig:CLmmax}
\end{figure}

Actually, there is no unique value of $m\mathrm{_{max}}$ for a total number of stars ${\cal N}$, but the possible values of $m\mathrm{_{max}}$ are distributed following the probability function

\begin{eqnarray}
\Phi_{m_\mathrm{max}}({m_\mathrm{max}}|{\cal N}) &=& {\cal N} \, \phi(m_\mathrm{max}) \,  p(m < m_\mathrm{max})^{{\cal N} -1} = \\
&=& {\cal N} \, \phi(m_\mathrm{max}) \, \left(\int_\mathrm{m_{low}}^{m_\mathrm{max}} \phi(m) \,\mathrm{d}m\right)^{{\cal N} -1}, 
\label{eq:pdfmmax}
\end{eqnarray}

\noindent as  deduced by \cite{gumbel,sornette,vanAlvada,OC05,MC08,PK08}, among others. 

In Fig.~\ref{fig:pdfmmax} we show the distribution $\Phi_{m_\mathrm{max}}({m_\mathrm{max}}|{\cal N})$ for different values of ${\cal N}$. The circle on each pdf corresponds to the position of the characteristic value $\hat{m}_\mathrm{max}$, which divides the pdf in two areas: the left one containing the 37\% of the probability and the right one containing the 63\% of the probability. We note that $\Phi_{m_\mathrm{max}}({m_\mathrm{max}}|{\cal N})$ is highly asymmetrical.
Given the shape of the distribution, it cannot be described only by their parameters (mean, variance, and so on); we must consider the whole distribution for any comparison with the observational data. This can be done in two ways, by a percentile analysis (analysis around the median)  and by a confidence interval analysis around the mode\footnote{The analyses based on the  parameters of the distribution, on the percentile,  and on confidence intervals around the mode are equivalent only in  the Gaussian case, where $1\sigma$ is almost equivalent to the percentile range $16-84$\% and the 68\% confidence interval.} (the maximum value of the distribution, which is related to the most common value obtained in a set of observations).

Figure~\ref{fig:percentmmax} shows a percentile analysis of the distribution. The figure also includes the position of the mean, mode, and characteristic values of the distribution for reference. The position of the mean, $\left<m_\mathrm{max}|{\cal N}\right>$, mostly falls between the 63\% and 84\% percentile, i.e., far from the median of the distribution. On the other hand, $\hat{m}_\mathrm{max}$ corresponds, as predicted, to the 37\% percentile. Finally, the mode of the distribution lies in the lowest percentile range. The figure also shows the ($m_\mathrm{max}$, $\cal N$) values compiled by \cite{WKB10}, in which $m_\mathrm{max}$ is determined from observations and $\cal N$ is inferred from star counting in a given mass range\footnote{Except in a few cases, \cite{WK04} and \cite{WKB10} obtain ${\cal N}$ by extrapolating to the full IMF range the number of stars $N_\mathrm{a}$ observed above a specified mass or within a specified mass range. Then, $\cal M$ is obtained by means of ${\cal M} = {\cal N} \times \left<m\right>$. We obtained the plotted $\cal N$ values by division of the $\cal M$ values quoted in their tables by $\left<m\right>$.}. It also shows the data from \cite{KM11}, who quote the observed masses of  individual stars of 14 young stellar groups in four different regions ($m_\mathrm{max}$, $\cal N$, and $\cal M$ were obtained from their tabulated data). We also show the corresponding $m_\mathrm{max}$ and $\cal N$ values of field stars in each region analyzed by \cite{KM11}, which are in agreement with the general trend of the correlation.

The confidence interval around the mode analysis takes into account the distribution shape and the range of probability of any region in the diagram. This is done by sorting the contributions to the probability in decreasing order and finding the $m_\mathrm{max}$ range that contains some specified amount of probability. Different confidence intervals are obtained by adding the sorted probabilities, taking into account their associated $m_\mathrm{max}$ values. This methodology is extensively used in the analysis of redshifts in photometric surveys \cite[see][~for more details]{alberto}. The situation is illustrated in Fig.~\ref{fig:CLmmax}, which includes the 90, 68, and 26\% confidence intervals.

\subsection{The pdf of $\cal N$ for a known $m_\mathrm{max}$, $\Phi_{\cal N}({\cal N} | m_\mathrm{max})$ }
\label{subsec:mmaxfix}

\begin{figure}
\resizebox{\hsize}{!}{\includegraphics{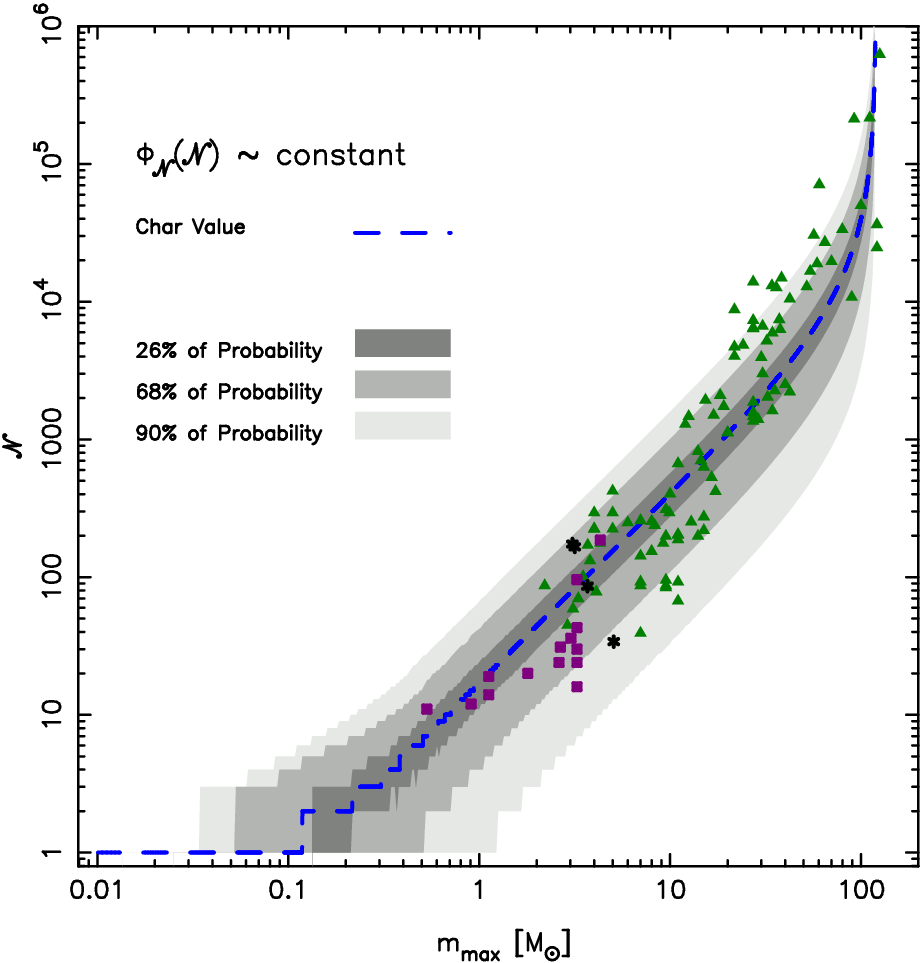}}
\caption[]{Confidence interval analysis of $\Phi_{\cal N}({\cal N}|m_\mathrm{max})$ as a function of $m_\mathrm{max}$ for a 
$\Phi_{\cal N}({\cal N}) = \mathrm{constant}$. Symbols have the same meaning as in Fig.~\ref{fig:percentmmax}.}
\label{fig:Nfuncmmax}
\end{figure} 

\begin{figure}
\resizebox{\hsize}{!}{\includegraphics{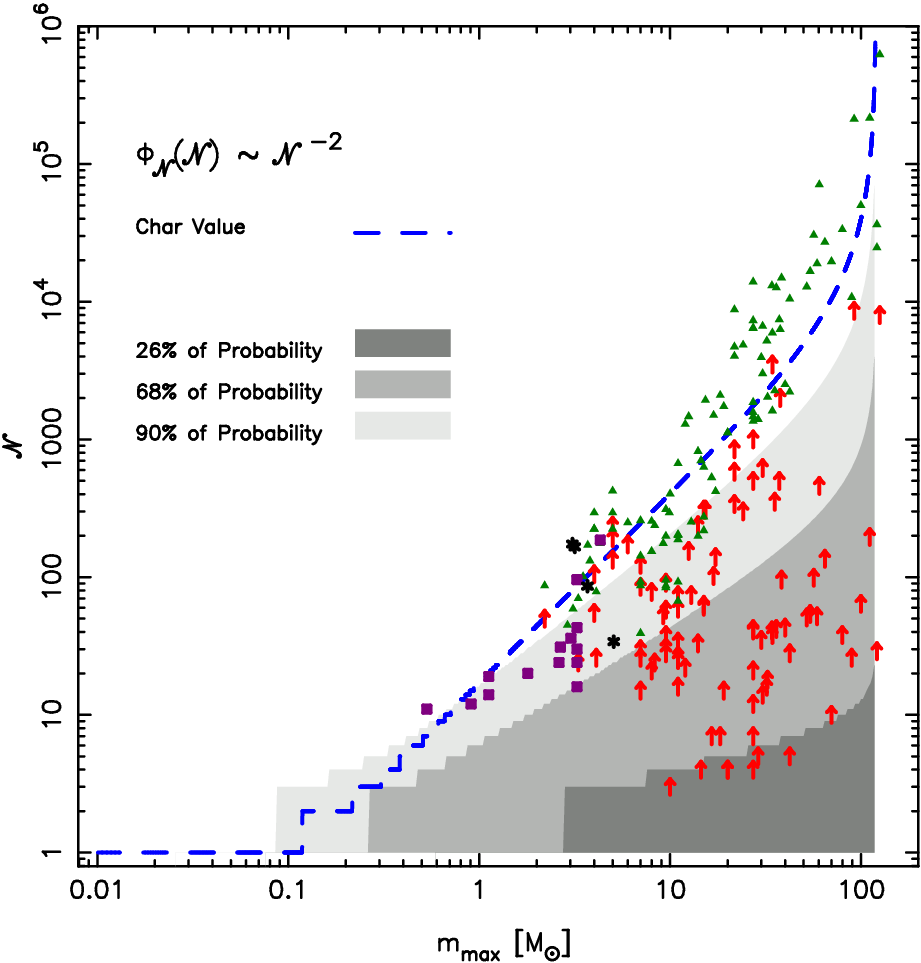}}
\caption[]{Confidence interval analysis of $\Phi_{\cal N}({\cal N}|m_\mathrm{max})$ as a function of $m_\mathrm{max}$ for a 
$\Phi_{\cal N}({\cal N}) \propto {\cal N}^{-2}$. Arrows: data points by \cite{WKB10} using ${\cal N}_\mathrm{obs}$ without correction of incompleteness due to unobserved stars. Other symbols have the same meaning as in Fig.~\ref{fig:percentmmax}.}
\label{fig:Nfuncmmax17}
\end{figure} 

In Sect.~\ref{subsec:Nfix} we discussed the estimation of $m_\mathrm{max}$, given the number of stars ${\cal N}$. 
Alternatively, we can also investigate the 
opposite case, the estimation of ${\cal N}$ from a known $m_\mathrm{max}$ (that is, the determination of the $\Phi({\cal N} | m_\mathrm{max})$ distribution). To address this problem, we can use the Bayes' theorem:

\begin{equation} 
\Phi_{\cal N}({\cal N} | m_\mathrm{max}) =  \frac{\Phi_{m_\mathrm{max}}({m_\mathrm{max}}|{\cal N}) \, \Phi_{{\cal N}}({\cal N})}{\int \Phi_{m_\mathrm{max}}({m_\mathrm{max}}|{\cal N})\; \, \Phi_{{\cal N}}({\cal N})\; \mathrm{d} {\cal N}}.
\end{equation}

We know all terms on the right-hand side of this equation, except $ \Phi_{{\cal N}}({\cal N})$,which is the probability of
having a system with a given total number of stars, i.e., an initial number-of-stars-per-cluster  function (an initial cluster number function, ICNF).
If  $\Phi_{{\cal N}}({\cal N})$ is a power-law distribution in a similar fashion to the initial cluster mass function (ICMF),
$\Phi_{{\cal N}}({\cal N}) = A {\cal N}^{-\beta}$ with $A$ a normalization value,  we find

\begin{equation} 
\Phi_{\cal N}({\cal N} | m_\mathrm{max}) =  A'\,  \,  p(m < m_\mathrm{max})^{{\cal N} -1}  \, {\cal N}^{1-\beta},
\label{eq:phiNmmax}
\end{equation}

\noindent where $A'$ is a normalization value that includes $A$.

The mode of  $\Phi_{\cal N}({\cal N} | m_\mathrm{max})$,  ${\cal N}^\mathrm{mode}$, 
is obtained by equaling to zero its first derivative with respect to ${\cal N}$, which yields\footnote{${\cal N}$Ê~is not a continuous variable; hence it cannot have been derivated and ${\cal N}^\mathrm{mode}$ must be an integer number. Thus, the formulae provide only an approximation.}

\begin{equation}
{\cal N}^\mathrm{mode} \approx \frac{\beta-1}{\ln  p(m < m_\mathrm{max})}.
\label{eq:Nmode}
\end{equation}

This equation has an acceptable solution only for $\beta < 1$; in particular, for a flat distribution of ${\cal N}$ (i.e., $\beta=0$) the result is approximately  
$1/p(m\geq m_\mathrm{max}$).  This justifies the name of $\hat{m}_\mathrm{max}$ as the {\it characteristic value},  since it provides $\cal N^\mathrm{mode}$ as a function of the most extreme value of the distribution under the hypothesis of a flat $\Phi_{{\cal N}}({\cal N})$\footnote{In Paper II we show that this 
assumption is implicit when $\cal N$ is inferred from the number $N_\mathrm{a}$ of massive stars in the ($m_\mathrm{max}$,  $m_\mathrm{a}$) range by using the relation ${\cal N} = N_\mathrm{a} \times p(m \geq m_\mathrm{a})$. Similarly, the assumption is implicit when $\cal M$ is inferred by multiplying the mean stellar mass by $\cal N$; it is a general assumption found in the literature and, in particular, is the method used  to infer $\cal M$ in  the \cite{WKB10} compilation.}. In Fig.~\ref{fig:Nfuncmmax} 
we plot the confidence intervals of the $\Phi_{\cal N}({\cal N} | m_\mathrm{max})$ distribution as a function of $m_\mathrm{max}$. We note that the axes of the plot have changed with respect to the figures in the previous section, since  $m_\mathrm{max}$ is now the variate. We also plot the data points from \cite{WKB10} and  \cite{KM11}.

However, Eq.~\ref{eq:Nmode} results in a negative value without astrophysical meaning if the ICNF is similar to the ICMF; $\Phi_{\cal N}({\cal N} | m_\mathrm{max})$ is a decreasing function for all $\cal N$, and the most probable ${\cal N}$ corresponds to the maximum of $\Phi_{{\cal N}}({\cal N})$, i.e., the lower limit of the $\Phi_{\cal N}({\cal N})$ distribution. Hence, $\Phi_{{\cal N}}({\cal N})$ modifies the confidence interval analysis of $\Phi_{\cal N}({\cal N} | m_\mathrm{max})$, as shown in  Fig.~\ref{fig:Nfuncmmax17}

It seems surprising that, depending the independent variable used ($m_\mathrm{max}$ or $\cal N$), one has to take into account $\Phi_{\cal N}({\cal N})$.
Where is the $\Phi_{\cal N}({\cal N})$ dependence in Figs.~\ref{fig:percentmmax} and \ref{fig:CLmmax}? Actually, we must be aware that  Figs.~\ref{fig:percentmmax}, \ref{fig:CLmmax}, \ref{fig:Nfuncmmax}, and \ref{fig:Nfuncmmax17} are not representations of $\Phi_{{\cal N},m_\mathrm{max}}({\cal N}, m_\mathrm{max})$, which would be the one to be compared with observational data. Instead, they are a representation of the probability {\it for fixed} values in the $x$-axis, i.e., the figures {\it can be only interpreted making vertical (discrete or infinitesimal) slices}. Hence, for comparison with data, the x-axis on Figs.~\ref{fig:percentmmax} and \ref{fig:CLmmax} must be weighted by $\Phi_{\cal N}({\cal N})$, and the 
x-axis on Figs.\ref{fig:Nfuncmmax} and \ref{fig:Nfuncmmax17} must be weighted by $\phi(m)$. Obviously, such a weight process changes the probability density in the ${\cal N} - m_\mathrm{max}$ plane.

\subsection{Which information does the ${\cal N} ~(\mathrm{or}~{\cal M})-m_\mathrm{max}$ plane contain?}

All the quantities considered here, $m_\mathrm{max}$, ${\cal N}$, and $\cal M$, have their own distributions, $\phi(m)$, $\Phi_{\cal N}({\cal N})$, and $\Phi_{\cal M}({\cal M})$. So, any uncertainty of data points in the ${\cal N} ~ (\mathrm{or}~{\cal M})-m_\mathrm{max}$ plane would be minimized or amplified by such distributions, and
 neither $\Phi_{m_\mathrm{max}}( m_\mathrm{max} | {\cal N} )$ nor $\Phi_{\cal N}({\cal N} | m_\mathrm{max})$ (or their $\cal M$ counterparts) are suitable descriptions. The only suitable distribution of data points is given by $\Phi_{m_\mathrm{max},{\cal N}}( m_\mathrm{max}, {\cal N} )$\footnote{That is: 
\begin{eqnarray}
\Phi_{m_\mathrm{max},{\cal N}}(m_\mathrm{max},{\cal N}) &=& \Phi_{m_\mathrm{max}}(m_\mathrm{max}|{\cal N}) \, \Phi_{\cal N}({\cal N})\nonumber\\
&=& \Phi_{\cal N}( {\cal N} | m_\mathrm{max} ) \, \phi(m_\mathrm{max})\nonumber
\end{eqnarray}} (or their $\cal M$ counterpart, see below). 
This pdf is shown in Fig.~\ref{fig:Nmmax} for the case of a $\Phi_{\cal N}({\cal N}) \propto {\cal N}^{-2}$.
However, the use of  $\Phi_{m_\mathrm{max},{\cal N}}( m_\mathrm{max}, {\cal N} )$  imposes some important caveats.

The first of these caveats affects any test on the ${\cal N} ~(\mathrm{or}~{\cal M})-m_\mathrm{max}$ correlation. Such a test can only be done at a distribution level and not  in a data-point-by-data-point analysis. This means  that we need a quantitative characterization of the uncertainty associated to each data point and must combine the corresponding uncertainties to obtain a density map in the ${\cal N} ~(\mathrm{or}~{\cal M})-m_\mathrm{max}$ plane.

The second caveat refers to the plane to be used:  ${\cal N} -m_\mathrm{max}$ or ${\cal M}-m_\mathrm{max}$? It includes two different aspects. The first is that any $\cal M$ inference implicitly includes an  $\cal N$ inference, and in most of the cases 
(all where $\left< m \right>$ is used), it is actually an $\cal N$ inference itself but expressed as $\left< {\cal M} \right>$ (i.e.,  the plane to be used is actually ${\cal N} -m_\mathrm{max}$).
The second aspect is that the distribution of data points in the ${\cal N} -m_\mathrm{max}$ plane includes $\phi(m)$ and  $\Phi_{\cal N}({\cal N})$ and the distribution of data points in the ${\cal M} -m_\mathrm{max}$ plane {\it also} includes $\Phi_{\cal M}({\cal M})$.
This means that some hypothesis about the relation between $\cal N$ and $\cal M$ is always required when the ${\cal M} -m_\mathrm{max}$ plane is used.

We conclude this section with a brief discussion about the falsification of the random sampling of the IMF claimed by \cite{WKB10} in view of the results presented here, that is, the dependence on $\Phi_{\cal N}({\cal N})$ and $\Phi_{\cal M}({\cal M})$ in the distribution of data points in the 
${\cal N} ~(\mathrm{or}~{\cal M})-m_\mathrm{max}$ plane.

First, random sampling is an axiom in statistics and probability. It is not a hypothesis.  Statistical tests evaluate the compatibility of a hypothetical distribution with a given sample. There can be two main reasons for the incompatibility of  both entities: 
(a) the assumed distributions are not a correct representation of the sample, (b) the sample is biased or not randomly chosen.
In the present case, the hypothesized distributions are the IMF, the ICNF, and the ICMF, where the ICMF and the ICNF are linked not trivially by Eq.\ref{eq:Mtot}. We would assume a universal IMF, but still need an ICMF (or ICNF) characterization. The very definition of the ICMF (or ICNF) leads to an uncomfortable situation similar to the case of the IMF:  we have no means of defining an empirical sample that can be directly related to SF theories without introducing a major assumption, that is, the cluster definition. Can a single star be considered as a valid cluster? How do we define a single cluster formation event in a giant molecular cloud? Is there a difference between the ICMF defined over a random set of clusters and the one defined over a group of clusters that would have a common origin in a large-scale star-forming event?

Hence, the results obtained by \cite{WKB10} can be interpreted in different ways:

\begin{itemize}
\item The clusters in the sample do not follow the assumed IMF. 
\item The clusters in the sample do not follow the assumptions about the ICMF or ICNF. 
\item The sample is biased due to selection effects (including the definition of what a cluster is).
\item The sample is incomplete, so no conclusions about the preceding items can be obtained.
\end{itemize}

We will discuss these issues in more detail in Papers II and III.

\begin{figure}
\resizebox{\hsize}{!}{\includegraphics{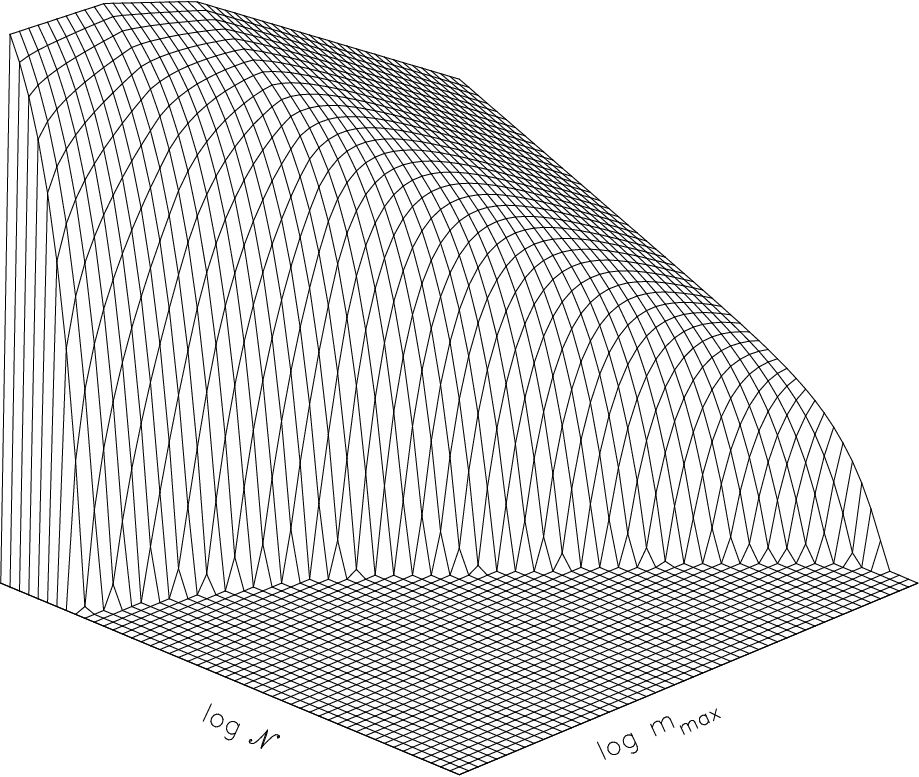}}
\caption[]{3D representation of $\log \Phi_{m_\mathrm{max},{\cal N}}(m_\mathrm{max}, {\cal N})$ distribution  for a 
$\Phi_{\cal N}({\cal N}) \propto {\cal N}^{-2}$.}
\label{fig:Nmmax}
\end{figure}

\section{Discussion}
\label{sec:discusion}

In the previous sections we have established the formal probabilistic interpretation of the IMF and the propagation of this interpretation in the 
correlation between $m_\mathrm{max}$ and $\cal N$. We can now explore the implications of such an interpretation and (a) compare it with the implications of concurrent interpretations (Sect.~\ref{sec:Mmmax}), and (b) discuss the random-sampling assumption of this work and its implications for the relation between the  IMF and the SF (Sect.~\ref{sec:sampling}).

\subsection{Literature on the ${\cal M}-m_\mathrm{max}$ and the ${\cal N}-m_\mathrm{max}$ correlations}
\label{sec:Mmmax}

There are copious studies related to the existence and modeling of a ${\cal M}-m_\mathrm{max}$ correlation
\cite[for instance, ][]{Red78,Lar82,Van82,GVD94,GVDB95,Elm97,Elm99,Elm00,Lar03,KW03,WK04,OC05,WK06,PG07,SM08,MC08,WKB10,Kroetal11}. 
Some of these articles give an explicit formulation of this relation, while others propose that it is a physical relation that links both quantities.
Others even argue that the relation is not physical but only an effect of the size of samples. As we will see, the difference
 among the various ${\cal M}-m_\mathrm{max}$ relationships  and their meaning does not depend on the relation itself, but rather on how each author interprets the IMF.

One common assumption is that the ${\cal N}-m_\mathrm{max}$ and the ${\cal M}-m_\mathrm{max}$ correlations are theoretically equivalent. With this idea in mind, the first correlation  is preferred by \cite{SM08} and \cite{MC08}, who argue that $\cal N$ is the natural independent variable for testing the random-sampling hypothesis. The second one is preferred by \cite{WKB10} because, with the two quantities inferred, the possible error in $\cal N$ is larger than the error in $\cal M$.  Only a few authors  \citep{SM08} explore the question of whether they are indeed formally equivalent or not. As we have seen previously, in a probabilistic framework they are not equivalent (cf., Eq.~\ref{eq:Mtot}).

\subsubsection{The IMF as an exact analytical law}

\begin{figure}
\resizebox{\hsize}{!}{\includegraphics{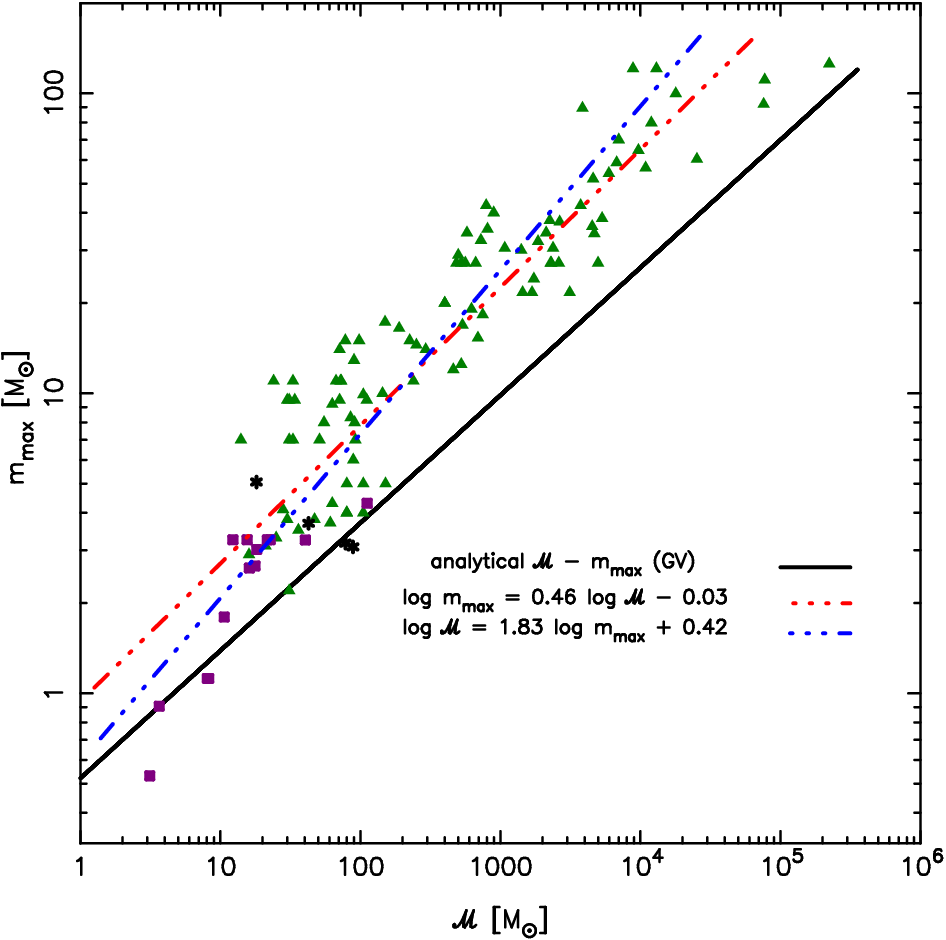}}
\caption[]{${\cal M}-m_\mathrm{max}$ relationship resulting from the analytical formulation of the IMF of \cite{GVD94,GVDB95}. The figure includes data points from \cite{WKB10} and \cite{KM11}, where symbols have the same meaning as in Fig.~\ref{fig:percentmmax}
and the result of two linear fits to the data from  \cite{WKB10} and \cite{KM11} using either $\log {\cal M}$ or $\log m_\mathrm{max}$ as the independent variable.}
\label{fig:MmaxplotGV}
\end{figure}

Let us consider the case of \cite{GVD94} and \cite{GVDB95} as an example of this interpretation. They assume that the IMF is not a probability distribution 
but an exact analytical law, $\phi_{\mathrm{GV}}(m)  = k({\cal M}) \,\times\, \phi(m)$, where  $k({\cal M})$ is a renormalization constant that, because ${\cal M}$ is the exact value of the amount of gas transformed into stars, verifies

\begin{equation}
{\cal M} =  \int_{\mathrm{m_{low}}}^{\mathrm{m_{up}}} m \, \phi_{\mathrm{GV}}(m) \, \mathrm{d}m = k({\cal M}) \int_{\mathrm{m_{low}}}^{\mathrm{m_{up}}} m \, \phi(m) \, \mathrm{d}m,
\label{eq:MGV}
\end{equation}

\noindent where $\phi(m)$ is the standard functional form of the IMF. The {\it exact} number of stars with mass $m_\mathrm{a}$ in the cluster is given by $N_\mathrm{a}=  \phi_{\mathrm{GV}}(m_\mathrm{a})$,
which implies that $k({\cal M})={\cal N}$.
Taking into account that stars are discrete entities, they propose a scenario in which
only the stellar masses that verify $\phi_{\mathrm{GV}}(m) \ge 1$ represent acceptable physical solutions (the so-called {\it richness effect}). Given that $\phi_\mathrm{GV}(m)$ decreases with $m$,  the most massive star in the cluster is the one that verifies

\begin{equation}
 \phi_{\mathrm{GV}}(m_\mathrm{max}) =  {\cal N} \times \phi(m_\mathrm{max})  = 1.
\label{eq:NGV}
\end{equation}

For a power-law IMF, $\phi(m) = A \, m^{-\alpha}$, this leads to a ${\cal M}-m_\mathrm{max}$ relationship with the form:

\begin{equation}
m_\mathrm{max} \propto {\cal M}^\frac{1}{\alpha}.
\label{eq:MmGV}
\end{equation}

According to the scenario proposed,  
the cluster forms stars in a sorted way, in which the stars with an associated larger value of $\phi_\mathrm{GV}(m)$ take precedence over stars with associated lower values of $\phi_\mathrm{GV}(m)$. So, the most massive star (the one with the lowest $\phi_\mathrm{GV}(m_\mathrm{max})$ value) is conditioned to the formation of a large enough number of lower mass star (the {\it richness effect}). Stated otherwise, the mass of this most massive star is determined by the amount of gas that remains after {\it all} possible lower mass stars have been formed with relative numbers established by the IMF. We note that the relevant point here is that  {\it there must be a certain amount of mass transformed into stars with mass $m < m_\mathrm{a}$  in order to have a star with mass  $m_\mathrm{a}$}. 

A similar  ${\cal M}_\mathrm{cloud} - m_\mathrm{max}$ relationship is found by \cite{Lar82,Lar03}. However, Larson's results come from fitting the observational data of {\it cloud} masses, ${\cal M}_\mathrm{cloud}$, with respect to $m_\mathrm{max}$, and they are quoted as a statistical correlation, not a physical law.
We note that a correlation between ${\cal M}_\mathrm{cloud}$ and $m_\mathrm{max}$ does not imply the same correlation between ${\cal M}$ and $m_\mathrm{max}$, since an efficiency factor is required
\cite[see][ for a more detailed discussion]{SE11}.
 
  In Fig.~\ref{fig:MmaxplotGV} we show the resulting ${\cal M}-m_\mathrm{max}$ relationship under these assumptions on the IMF and assuming the  functional form of the IMF used in this work. The figure includes data points from \cite{WKB10} and \cite{KM11}. We have included the result of two linear fits to the data from \cite{WKB10} and \cite{KM11} using either $\log {\cal M}$ or $\log m_\mathrm{max}$ as the independent variable. The theoretical relation is off toward larger  $\log {\cal M}$ values.

This interpretation of the IMF stems from stellar counting procedures. Since $\phi_{\mathrm{GV}}(m)$ is a continuous function, it cannot return a natural number $N_\mathrm{a}$ for any mass value $m_\mathrm{a}$; because  stars are discrete entities, this approach can only be an approximate description. This alone is sufficient to invalidate Eq.~\ref{eq:NGV} as a way to obtain the actual most massive star, since $\cal N$ may (unphysically) turn out to be a non-natural number. A consequence, this equation can only provide an approximation. 

This situation implies that {\it continuous functional forms of the IMF can only be directly related to the number of stars with a given mass interval}, and not to the number of stars with a given mass. This possibility is explored in the next interpretation case.

\subsubsection{The IMF as a distribution of the number of stars}
\label{subsubsec:IMFfrecuency}

One alternative view of the IMF is that it can be arbitrarily normalized and provide the exact number of stars  in a given {\it mass range}. This is the case assumed by \cite{Red78,Van82,Elm97,Elm99, Elm00,KW03,WK04,Elm06,WK06,WKB10} and \cite{Kroetal11}.
We refer to these articles as those  that use  the IMF de facto as a distribution of the number of stars.
Their interpretation is that the number of stars between $m_\mathrm{a}$ and $m_\mathrm{b}$ , with $m_\mathrm{a} < m_\mathrm{b}$, is given by

\begin{equation}
N (m \in [m_\mathrm{a},m_\mathrm{b}])=  \int_{m_\mathrm{a}}^{m_\mathrm{b}}  \, \phi_{\mathrm{Elm}}(m) \, \mathrm{d}m,
\label{eq:NElm}
\end{equation}

\noindent where $\phi_{\mathrm{Elm}}(m) = k \times \phi(m)$ with $k$ a normalization constant. This equation is the general case of Eq.~\ref{eq:charval}, that is, the definition of $\hat{m}_\mathrm{max}$, described above. The difference with the previous case is that the total number of stars in the cluster is now given by

\begin{equation}
{\cal N} =  \int_{\mathrm{m_{low}}}^{\mathrm{m_{up}}}   \phi_{\mathrm{Elm}}(m) \, \mathrm{d}m,
\label{eq:NtotElm}
\end{equation}

\noindent so, $k={\cal N}$. The actual total mass is given by integration of $m\,\times\,  \phi_{\mathrm{Elm}}(m)$  within the same mass limits. However, how the limits are written and what interpretation is given to them varies according to the author. Here we use the formalization by \cite{Elm97,Elm99,Elm00,Elm06}:

\begin{equation}
{\cal M} =  \int_{\mathrm{m_{low}}}^{\mathrm{m_{up}}} m \, \phi_{\mathrm{Elm}}(m) \, \mathrm{d}m = {\cal N} \times \int_{\mathrm{m_{low}}}^{\mathrm{m_{up}}} m \, \phi(m) \, \mathrm{d}m,
\label{eq:MElm}
\end{equation}

\noindent and postpone to the next subsubsection the discussion of the special case of \cite{WK04,WK06}, \cite{WKB10}, and \cite{Kroetal11}. Whatever the normalization is, we need an additional assumption to obtain the actual maximum stellar mass in the cluster from Eq.~\ref{eq:NElm}. 
We have to assume ad hoc that the most massive star $m_\mathrm{max}$ is the result of solving Eq.~\ref{eq:charval} (i.e., that $\hat{m}_\mathrm{max}$ is  the actual $m_\mathrm{max}$). To do so, external arguments, similar to the   {\it richness effect}, are required.
 
For a power-law IMF and $\mathrm{m_{up}} = \infty$, the  $m_\mathrm{max}-{\cal M}$ correlation is

\begin{equation}
m_\mathrm{max} \propto {\cal M}^{\frac{1}{\alpha-1}}  \propto {\cal N}^{\frac{1}{\alpha-1}}.
\end{equation}

\cite{Elm97,Elm99,Elm00} argue that, since the cluster is filled through random sampling, the inferred $m_\mathrm{max}$ can only be an estimate of the actual value. Only \cite{Van82}  states that it is possible to obtain the actual $m_\mathrm{max}$ value.

\begin{figure}
\resizebox{\hsize}{!}{\includegraphics{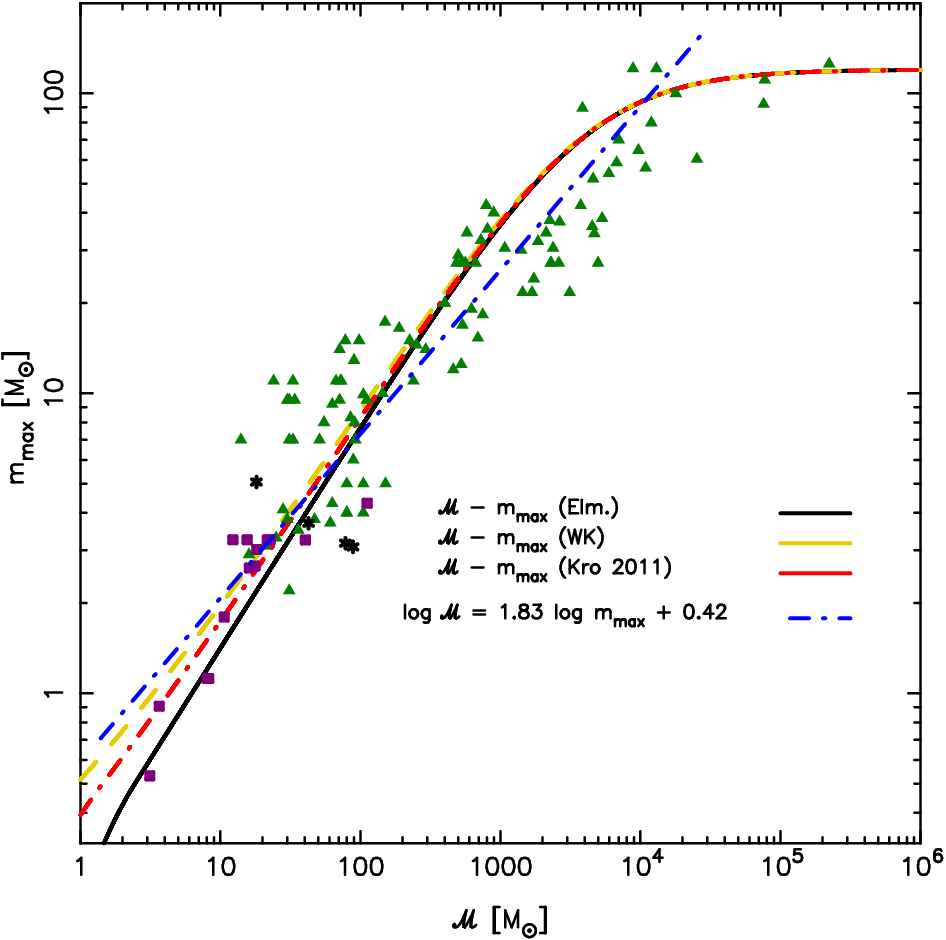}}
\caption[]{${\cal M}-m_\mathrm{max}$ relationship resulting from the distribution function formulation of the IMF of \cite{Elm97,Elm99,Elm00}, the formulation of \cite{WK04,WK06}, and the optimal sampling formulation of \cite{Kroetal11}. 
The figure includes data points from \cite{WKB10} and \cite{KM11} and the result of the linear fit of the data to $\log {\cal M}$ as a function of $\log m_\mathrm{max}$.}
\label{fig:Mmaxplot}
\end{figure}

In Fig.~\ref{fig:Mmaxplot} we show the resulting ${\cal M}-m_\mathrm{max}$ correlation under these assumptions using the functional form of the IMF employed here. The curve is completely equivalent to the $\left<{\cal M}\right> - \hat{m}_\mathrm{max}$ correlation obtained in the pdf case. The figure includes data points from \cite{WKB10} and \cite{KM11} just for comparison. We also included the result of a linear fit  of $\log {\cal M}$ as a function of $\log m_\mathrm{max}$ obtained from the data.

This interpretation of the IMF relies on stellar counting followed by a binning process. It is by far the most common interpretation and is assumed in a wide range of situations, from IMF determinations to stellar population synthesis. Its main feature is that Eq.~\ref{eq:NElm} provides the {\it actual}Ê number of stars and that ${\cal M} = {\cal N} \times \left< m \right>$ provides the {\it actual} total stellar mass in the cluster (this last feature is also 
shared by the analytical law interpretation). In this case it may seem that the 
problem with integer numbers of stars mentioned in the previous case is solved as far as we can always 
choose a suitable set of bins such that Eq.~\ref{eq:NElm} produce a natural number for any $m_\mathrm{a}$ and $m_\mathrm{b}$ values. However, the solution is 
not so trivial: depending on the bin definition, distributions with different shapes are obtained \citep{DS86,MAU05}, but the shape of the IMF is still defined by  $
{\cal N} \times \phi(m)$. Consequently, the bins cannot be defined at will. The only plausible solution is to assume that Eq.~\ref{eq:NElm} (and hence Eq.~\ref{eq:MElm}) is only valid in the limiting case ${\cal N} = \infty$ \cite[][]{CVGLMH02,FL10,Pisetal11}, and that, for finite $\cal N$ values, they do not provide actual $N (m \in [m_\mathrm{a},m_\mathrm{b}])$ or $\cal M$ values but only  
{\it estimates} of such values.
Again, we must understand what exactly this estimate represents.

To summarize this section, {\it no continuous functional form of the IMF can provide the actual number of stars, neither for a given mass nor for a given mass interval, but only an  estimate of it}. The only way to give  meaning to this estimate is by adopting a  probabilistic framework. This implies using a probabilistic algebra, which explicitly prevents arbitrary normalizations of $\phi(m)$.

\subsubsection{The Weidner \& Kroupa case}
\label{subsec:WK}

The studies by \cite{WK04,WK06,WKB10}, and \cite{Kroetal11} are another example of an interpretation of the IMF in terms of a distribution of the number of stars. However, they deserve  special attention since they represent a major effort to include conditions in the IMF.

The equations  to find a  ${\cal M}-m_\mathrm{max}$ relationship proposed by \cite{WK04,WK06}, once corrected by an improper account of $m_\mathrm{max}$ in $\cal M$ \cite[][]{Kroetal11}, are

\begin{eqnarray}
1   & = & \int_{m_\mathrm{max}}^{\mathrm{m_{up}}}  \phi_{\mathrm{WK}}(m) \, \mathrm{d}m,
\label{eq:mmaxWK}\\
{\cal M} - m_\mathrm{max} & =&  \int_{\mathrm{m_{low}}}^{m_\mathrm{max}} m\,  \phi_{\mathrm{WK}}(m) \, \mathrm{d}m \label{eq:MclWK}.
\end{eqnarray}

As in the previous case,  Eq.~\ref{eq:mmaxWK} is equivalent to the definition of $\hat{m}_\mathrm{max}$ given in Eq.~\ref{eq:charval} and  $\phi_\mathrm{WK}(m)$ has the same functional form (scaled by a constant $k_\mathrm{WK}$). A simple inspection shows that $k_\mathrm{WK}={\cal N}$. The difference with the previous case is in Eq.~\ref{eq:MclWK}:  the upper limit of the integral is $m_\mathrm{max}$ and not $m_\mathrm{up}$.  By doing so, \cite{Kroetal11} aim to constrain the IMF in such a way that Eq.~\ref{eq:mmaxWK} provides the {\it actual} $m_\mathrm{max}$ value rather than an {\it estimate} of it. 

They justify that Eq.~\ref{eq:mmaxWK} provides such actual value by focusing on how the IMF is sampled.
Their first approach was the sorted sampling scenario \cite[][]{WK06}, according to which the IMF is sort-sampled, where the stars with the lowest mass are those that form first. This scenario is physically motivated, based on the hydrodynamical simulations of cluster formation in 
competitive accretion without the inclusion of possible (positive or negative) feedback of massive stars \cite[][]{Bonetal03, Bonetal04}. \cite{WK06} presented Monte Carlo simulations to support this model, where clusters with a given total mass ${\cal M}$ are drawn from a randomly sampled IMF. The number of stars used in the simulation was estimated from ${\cal M}$ divided by the mean stellar mass. After that, the sample is sorted and the desired ${\cal M}$ value approximated  by 
accepting or rejecting the most massive star in the cluster. The most recent work  \cite[][]{Kroetal11} is based on the concept of the {\it optimal sample}: sampling is optimal if Eq.~\ref{eq:mmaxWK} is verified and produces the actual value of $m_\mathrm{max}$. In both cases, it is argued that the IMF is not random sampled. Figure~\ref{fig:Mmaxplot} shows the original and the corrected ${\cal M}-m_\mathrm{max}$ relationship they obtain.

This interpretation is based on a strict vision of the IMF as a stellar counting process involving an individual star,  the one with $m = m_\mathrm{max}$, and a stellar counting plus binning procedure for the remaining ${\cal N} -1$ stars. This can be seen from the treatment of the integral limits or equivalently, the histograms bins, throughout the different versions. In the original set of equations proposed by \cite{WK06}, $m_\mathrm{max}$ was counted twice in two non-overlapping bins. The new version \cite[][]{Kroetal11} clearly states the bin where $m_\mathrm{max}$ is, but now it opens a problem with the $\phi(m)$ definition. We recall that it is mainly a problem of inclusion of conditions, which is not a trivial issue. Let us consider the possible self-consistent cases:

\begin{enumerate}

\item 
We use the criteria of {\it equal to or larger than} for lower integral limits and {\it lower than} for upper ones  to give a physical meaning to  
Eq.~\ref{eq:mmaxWK}. However, if we want $m_\mathrm{max}$ to appear directly in the computation of $\cal M$, we must impose it ad hoc, which is done by using ${\cal M} - m_\mathrm{max}$ instead of $\cal M$. A self-consistent formulation, taking into account the integral limits in  Eq.~\ref{eq:mmaxWK}, is  to write explicitly the mass contribution of the stars in the ($m_\mathrm{max}$, $\mathrm{m_{up}}$) range

\begin{eqnarray}
m_\mathrm{max} & = & \int^{\mathrm{m_{up}}}_{m_\mathrm{max}} m\,  \phi_{\mathrm{WK}}(m)\,\,\,\, \Rightarrow \nonumber\\
\phi_{\mathrm{WK}}(m) &=& \delta(m-m_\mathrm{max}) + ({\cal N}Ê- 1) \times \phi(m| m < m_\mathrm{max}),  \label{eq:WKcor1}
\end{eqnarray}

\noindent where $\delta(m-m_\mathrm{max})$ is the Dirac delta function. However, this implies an ad hoc variation of the $\phi(m)$ functional form,
which is necessary to impose that $m_\mathrm{max}$ is the maximum stellar mass.

\item We use the criteria of {\it larger than} for lower integral limits and {\it equal or lower than} for upper ones. Then, we can compute $\cal M$ properly using $m_\mathrm{max}$ as the upper integral limit.  However, in this case we must change Eq.~\ref{eq:mmaxWK} by

\begin{eqnarray} 
0 & = & \int_{m_\mathrm{max}}^{\mathrm{m_{up}}} \phi_\mathrm{WK}(m) \mathrm{d}m \nonumber\\
&& \Rightarrow \phi_{\mathrm{WK}}(m) = k_\mathrm{WK} \,\times\,\phi(m|m\leq m_\mathrm{max}),  \label{eq:WKcor2}
\label{eq:mmaxalter}
\end{eqnarray}

\noindent which means that there is no star more massive than $m_\mathrm{max}$. This means, however, that we lose the equation giving $m_\mathrm{max}$ value, which must be imposed ad hoc.

\end{enumerate}

Cases (1) and (2) above are the only possible ones, and both constrain ad hoc $m_\mathrm{max}$ to be the maximum stellar mass in the cluster. Now, we have shown previously that any description of the IMF as a continuous function implicitly eliminates the dependence with $\cal N$ (and hence $\cal M$) and its interpretation as a distribution by number. The \cite{Kroetal11} case clearly shows that there is no way to include constraints into a distribution-by-number description of the IMF and, at the same time, enjoy the advantages of a continuous distribution representation. 
Once a continuous functional form for $\phi(m)$ is assumed, only a pdf interpretation is valid, and we implicitly renounce obtaining actual values of stellar masses, actual total masses, or actual values of $m_\mathrm{max}$. In particular, it would not be possible to obtain a hidden physical law implicit in the $\phi(m)$ functional form. At most we could obtain statistical correlations like the $\left<{\cal M}\right> - \hat{m}_\mathrm{max}$. If there were such physical laws, their origin would be external to the IMF and could only be inferred from detailed simulations, and not from algebraic manipulation of the IMF.
That is the price we must pay for the advantages of a continuous formulation of the IMF.

\subsubsection{The probabilistic  case}
\label{subsubsect:probcase}

The IMF is treated as a probability distribution in \cite{OC05,Elm06,PG07,MC08,SM08,HA10}, among others. Their basic assumption is similar to the one of this paper, and some partial results of the description shown here have been obtained by other authors \cite[including ][]{WKB10}. Here, we summarize the results from works on the topic in the global context of the formulation given in the previous section.
The common point of these works is that,  without additional {\it ad hoc}Ê~conditions, an ${\cal M}-m_\mathrm{max}$ relationship cannot be defined trivially as a physical law, but only as a statistical correlation. The total mass in the cluster, the total number of stars in the cluster, and the particular number of stars with given stellar masses are not fixed quantities, but  distributed ones, and none of them can be obtained univocally from the others. 
Hence, the use of ${\cal M}-m_\mathrm{max}$ or the use of ${\cal N}-m_\mathrm{max}$ is not just a question of  choice in terms of observational considerations;  it is actually the result of  statistical correlations of different distributions.

The probabilistic description of the IMF is included, by construction, in works that make use of Monte Carlo simulations \cite[see][~as examples]{WK06,Elm06,PG07,SM08,HA10}, where the IMF is sampled star by star up to a given value of $\cal M$ or $\cal N$. Such Monte Carlo simulations have been devoted to explain and compare different results using  different sampling algorithms.  \cite{HA10} made an explicit, exhaustive, and detailed study  of the issue.  As far as we know, only \cite{Elm06} and \cite{SM08} have made  theoretical studies aimed of describing  the relationship of  ${\cal M}-m_\mathrm{max}$ using conditional probabilities. 

Most of the theoretical studies have been carried out in terms of an ${\cal N}-m_\mathrm{max}$ relationship, using $\cal N$ as variate and $m_\mathrm{max}$ as variable and making use of  $\Phi_{m_\mathrm{max}}({m_\mathrm{max}}|{\cal N})$. They often  include an expression for the mean value of the distribution \citep{OC05},  the mode of the distribution \citep{gumbel,KS04}, or the percentile analysis \citep{WKB10}. However, there is almost no study in terms of the $m_\mathrm{max}-{\cal N}$ relationship nor in the $\Phi_{\cal N}({\cal N})$ dependence of the ${\cal N}-m_\mathrm{max}$ correlation \citep{Elm06,SM08}.

So, in the probabilistic case, the  ${\cal N}-m_\mathrm{max}$, ${\cal M}-m_\mathrm{max}$, $m_\mathrm{max}-{\cal N}$, and $m_\mathrm{max}-{\cal M}$ correlations are {\it not} equivalent to each other. The ${\cal M}-m_\mathrm{max}$ correlation requires a $\Phi_{\cal N} ({\cal N}|\cal M)$ distribution which is not required by the ${\cal N}-m_\mathrm{max}$ correlation. In addition, establishing the $m_\mathrm{max}-{\cal N}$ and $m_\mathrm{max}-{\cal M}$ correlations requires some priors about the distribution of $\Phi_{\cal N} ({\cal N})$ and $\Phi_{\cal M} ({\cal M})$ that are not considered in the previous correlations.

The probabilistic formulation offers the advantages of using continuous distributions and including conditions formally. However, this does not mean that any condition can be represented analytically.
We have mentioned above that the \cite{WK04,WK06} formulation is a major effort to include conditions in the IMF. Let us rewrite Eq.~\ref{eq:WKcor1} in statistical terms and give a meaning to such distribution:

\begin{eqnarray}
\phi(m|m_\mathrm{max};{\cal N}) & =&  \frac{\delta(m-m_\mathrm{max})}{\cal N} + \frac{{\cal N} - 1}{\cal{N}} \phi(m| m < m_\mathrm{max}).
\end{eqnarray}

The above equation describes the constrained IMF {\it for a fixed}Ê $m_\mathrm{max}$ value in a set of $\cal N$ stars. This constraint does not imply that a star with $m_\mathrm{max}$ is present in the cluster, but just that there are no stars more massive than $m_\mathrm{max}$ and that the event $m=m_\mathrm{max}$ has a probability of $1/{\cal N}$. Since all the arguments of the characteristic value hold here, the associated  characteristic value is the fixed $m_\mathrm{max}$ value, which is also a cut-off value of the distribution. So, 63\% of realizations for clusters with $\cal N$ stars following such pdf have at least one star with mass $m_\mathrm{max}$ (and no stars more massive than $m_\mathrm{max}$).

Hence, there is no way to include in an analytical form the condition that the most massive star is actually $m_\mathrm{max}$ and that such a star is present in {\it any realization}. There is also a similar problem with ${\cal M}$, although the problem in this case is more severe since it also requires a $\Phi({\cal N})$ (discrete) distribution.
However, there is an infinite number of combinations of stellar masses that are consistent with any reasonable ${\cal M} - m_\mathrm{max}$ physical law.

The only possible solution at the moment to include a ${\cal M} - m_\mathrm{max}$ physical law and work with it is
to perform a large set of Monte Carlo simulations, which should assume a particular $\Phi({\cal N})$ distribution,
and just consider the subset where the chosen ${\cal M} - m_\mathrm{max}$ physical law is verified. Then, any physical result must be obtained  numerically (as opposed to analytically). The advantages of describing $\phi(m)$ as a continuous distribution are thus lost\footnote{We note that any sampling proposal that aims to reproduce a ${\cal M} - m_\mathrm{max}$ physical law with a finite number of stars $\cal N$ is also doomed to this situation: it provides a $\phi(m_i)$ array, but not a continuous $\phi(m)$ distribution.}.

\subsection{Sampling, iid variables,  and the relation of the IMF with SF}
\label{sec:sampling}

We have seen that the existence of a physical law linking ${\cal M}$ and $m_\mathrm{max}$ cannot be established through a simple manipulation of the IMF functional form.
he current debate on whether the IMF is randomly or non-randomly sampled stems mainly from works by \cite{WK06} and \cite{WKB10}, where $\hat{m}_\mathrm{max}$ is interpreted as the exact value of the most massive star in a cluster with a given mass. This debate has been focusing on different sampling proposals.
Even if the authors themselves now consider the sorted sampling proposal just as a first approximation \cite[][]{Kroetal11}, we want to emphasize that the key point of different sampling algorithms is not the sorting process, but the assumed relation between $\cal N$ and $\cal M$ (e.g., the sorted sampling proposal uses an $\cal N$ value estimated by means of $\cal M$ divided by $\left< m | m <  m_\mathrm{max}\right>$, which imposes a constraint in $\cal N$). The situation is actually more clearly described in  the {\it richness effect} proposed by \cite{GVD94,GVDB95}: a star with mass $m_\mathrm{a}$ is formed according to the amount of gas that remains in the system once a certain number of stars with $m < m_\mathrm{a}$ have been formed. The sampling problem appears when we
try to fix ${\cal M}(m < m_\mathrm{a})$ and ${\cal N}(m < m_\mathrm{a})$ simultaneously and include it analytically in the $\phi(m)$ functional form.

As we have shown, there is no self-consistent way to do it with the current description of $\phi(m)$.  The inclusion of any ${\cal M} - m_\mathrm{max}$ physical law, no matter what its interpretation is, precludes using an analytical functional form for the IMF.
The sampling methods proposed by different authors are actually operational methods, not an implementation of the physical process \footnote{The optimal sampling algorithm  provided by \cite{Kroetal11} is based on obtaining bins through the  {\it larger than} for lower integral limits and {\it equal to or lower than} for upper integral limits. These criteria are complementary to those
underlying their equations to obtain the ${\cal M} - m_\mathrm{max}$ relationship. In addition, the IMF is filled from $m_\mathrm{max}$ down to lower masses, contrary to the physical arguments given to justify the sorting sampling algorithm. We stress that it is not a problem of the formulation in as much as the physical formulation of the problem is not linked with the operational mathematical method used to solve the physical equations.}. 

However, we want to stress that the question on whether the IMF is randomly sampled or not
(i.e., whether stars are iids or not) is completely valid, independent of the particular problem motivating the question. So we will not attempt to discuss this question in terms of any specific results from literature, but from a more general perspective.

\subsubsection{Identical and independent distributed variables and the relation of the IMF with the star formation}

The question we aim to answer is: are stellar masses iid variables, or, at least, can they be treated as if they were? A sample is an iid sample if each random variable 
has the same identical probability distribution and all of them are mutually independent.  

Throughout the paper, we have explicitly excluded a mention to the SF physics. It is now time to take a look at different ways in which the SF and the IMF can be linked and how randomness enters in this game. There are several possible ways. (a) Some physicists prefer to assume a deterministic universe in which one and only one result is obtained for a given set of initial conditions. But there is such a large variety of initial conditions that they can be only described in a probabilistic way. Hence the results of SF events, like 
the IMF itself, can be only described in a probabilistic way.  (b) We can also assume an universe where determinism, although it exists, is somehow hidden by complexity. Thus we assume accordingly that the SF is a complex process in the mathematical sense: nonlinear and with interconnected components, producing such a large variety of results that they can only be treated in a probabilistic way. (c) We admit that there are intrinsically random variables in nature and that the SF is an intrinsically random process (like turbulence), so its results can only be treated in a probabilistic way. We refer to \cite{SE11,Sanetal06,Elm99,Elm11} as examples where some of these different scenarios are considered.

The feature  common to these three cases is that the IMF should be used probabilistically (i.e., stellar masses are randomly sampled), which does not imply that the SF is random. There would be no physical $\cal M$ and $m_\mathrm{max}$ relationship at all, or there would be a deterministic physical law linking $\cal M$ and $m_\mathrm{max}$. However, the internal distribution of stellar masses that are physically compatible (in the SF sense) with this physical law would depend on a set of unknown (and variable) initial conditions or intrinsically random characteristics. Then the IMF could only be described by means of a probabilistic formulation. A probabilistic interpretation of the IMF does not contradict a deterministic vision of the physics of SF. 

On a large scale, the IMF is the result of all possible SF events and SF modes, although it does not necessarily describe any particular one.  
Following this argument, we are able to describe probabilistically the incidence of having a star with a given mass that was born at a a given time, the stellar birth rate ${\cal B}(m,t)$, as the composition of two independent functions: the star formation history, SFH $\psi(t,{\cal M})$ (although $\psi(t,{\cal N)}$ would be more adequate) and the IMF, $\phi(m)$ \cite[][]{Sch59,Sch63,Tins80,Sca86}. The first function includes all the possible SF modes and provides the time-scale and the amount of gas transformed into stars. The second one describes how a given amount of gas would be distributed among different stellar masses. We recall that the first IMF determinations were done with field stars \cite[][]{Sal55}, so they implicitly averaged a large variety of SF modes.

The separation of  ${\cal B}(m,t)$ into two independent functions seems to be a valid approach for the study of galaxies and a variety of systems where different modes of star formation coexist; it has been extensively used in extragalactic astronomy and cosmology. One particular characteristic of this approach is the use of single stellar populations \cite[SSP,][]{RB86} which corresponds to $\psi(t,{\cal N}) = {\cal N} \times \delta(t)$. Since any function can be described by a sum of $\delta(t-\tau)$ functions, it allows  the SFH to be recovered from observational data or the evolution of galaxies to be described as a composition of SSPs with different intensity. The star formation rate, SFR, can then be defined as a time average of the SFH \cite[][]{dSFK12} or as the result of a flat SFH ($\psi(t,{\cal N}) = \mathrm{constant}$). Current SF rate indicators are based on SSP modeling with constant SFH \cite[][]{Ken88}.

The case would be different if we changed the scale to smaller systems. When we restrict the situation to specific SF modes, particular details emerge and have some imprint on the IMF. The more restrictive the mode, the more details are present. In this case we are moving ourselves to particular IMF realizations with given conditions, which may depart from the probabilistic description given by $\phi(m)$. At small scales, the validity of the decomposition of ${\cal B}(m,t)$ in two independent functions  is not clear. However, the universality of the IMF even at such scales leads one to think that it would be the case (however, see \citealt{Elm11} for an example of possible variations of the IMF, especially in the low-mass tail,  depending on the environmental conditions).

The approach we have presented here when talking about ${\cal B}(m,t)$ is a top-down  one: $\phi(m)$ is the most generic representation, so that the larger the system, the more valid it is.
We note that this vision is mentioned by \cite{Van82}, who also claimed existence of a ${\cal M}-m_\mathrm{max}$ physical law. Because there is an universal IMF at a large scale, he says, the IMF varies at small scale.

In this case it is expected the IMF has a quasi universal shape at high scales with possible variations at small scales. Here, we understand that deviations from a universal shape are allowed as far as they are small compared to the global budget. In addition, the incidence of deviations also depends on the size of the system, that is, the integral of the $\psi(t,{\cal N)}$ over time \cite[see][ for a discussion]{dSFK12}.

There is also a bottom-up approach when talking about ${\cal B}(m,t)$, which is the one proposed by the IGIMF theory. In this case, universality in the IMF functional form is assumed. However, there is a ${\cal M}-m_\mathrm{max}$ physical law that relates $\cal M$ with $m_\mathrm{max}$; hence there is IMF variability in the sense of a variable $m_\mathrm{max}$ for given $\cal M$. It is assumed that this physical law 
operates for all SF modes, or equivalently, that there is one SF mode: star formation in clusters. In this case, the mass distribution of stars depends on where (and when) they were formed, so only stars formed 
in the same cluster (or clusters with the same $\cal M$) share the same IMF.

For the study of galaxies or, in general, systems that may contain clusters with different masses, it is necessary to take into account the distribution of the total masses of these clusters: the ICMF. As a result, at a galactic scale there is not one IMF, but a IGIMF that results through the combination of the ICMF and different IMFs. It depends on $\cal M$ and implies a redefinition of the IMF itself \cite[][]{KW03}. In this case it is not clear if ${\cal B}(m,t)$ can be  separated into independent functions and how \cite[][]{Ceretal11}. This implies major revisions of global galactic and extragalactic studies,
including the SSP concept, 
and there is currently a large debate on the issue \cite[][]{Coretal09,Furetal11,Eld12}. Although a full discussion goes beyond the scope of this paper, we want to point out that there would be a $\left<{\cal M}\right> - \hat{m}_\mathrm{max}$ physical law, although it must be imposed {\it ad hoc}, and that, whatever the  case, random sampling and a probabilistic description of the IMF are compatible with it.

\section{Conclusions}
\label{sec:conclusions}

Having carried out a thorough analysis of different IMF interpretations, with a focus on the question of how information on $m_\mathrm{max}$ can be extracted from the IMF itself, we are in position to formulate the problem in a different way:
{\it What information does the IMF contain? Can we extract information on the SF process from an algebraic manipulation of the IMF?}
The answers to these questions are driven by the interpretation of the IMF adopted by each author and, in particular, their conclusion as to whether, without direct observations, $m_\mathrm{max}$ can be  exactly determined or just estimated.

Our analysis of the problem has led us to the following main conclusion:
 Only a probabilistic interpretation of the IMF, where $\phi(m)$ is a pdf (ruling out arbitrary normalizations) and stellar masses are random sampledly iid variables, provides a physical and mathematical self-consistent formulation that explains the $\left<{\cal M} \right> - \hat{m}_\mathrm{max}$ statistical correlation obtained from IMF algebraic manipulation. We also give plausible arguments that introduce the IMF as a probabilistic distribution when related with the physics of the star formation process.

Additional conclusions of this work are:

\begin{enumerate}

\item The actual total stellar mass of a cluster, $\cal M$, cannot be inferred from an IMF, $\phi(m)$, with a continuous functional form. A direct IMF integration only provides its mean value, $\left<{\cal M}\right>$, for a given number of stars $\cal N$:

\begin{equation}
 \left< {\cal M}\right> = {\cal N} \times \left< m \right> = {\cal N} \times \int_{\mathrm{m_{low}}}^{\mathrm{m_{up}}} \, m \, \phi(m) \, \mathrm{d}m.
 \label{eq:fin0}
 \end{equation}

Although some authors do not consider $\cal N$ as a relevant physical variable \cite[][]{Kroetal11},
the fact that stars are discrete entities and $\cal N$ is a natural number are
relevant physical constraints that must be included in the treatment of the 
IMF and in the algebra used to obtain physical results from it.

\item Given the equation defining the most massive star in a system,

\begin{equation}
\frac{1}{\cal N}   =  \int_{\hat{m}_\mathrm{max}}^{\mathrm{m_{up}}}  \phi(m) \, \mathrm{d}m, \nonumber
\label{eq:last}
\end{equation}

the resulting $\left<{\cal M}\right> -  \hat{m}_\mathrm{max}$ is practically independent of the specific IMF interpretation adopted. However, how this equation is understood strongly depends on the  framework of the interpretation. 

\item In a probabilistic interpretation, Eq.~\ref{eq:last} provides a {\it characteristic mass}, $\hat{m}_\mathrm{max}$, that is, the value of $m$  that is not reached or exceeded 
with a probability 0.37 in a sample of ${\cal N}$ stars, but not the actual mass of the most massive star in the sample. 

\item For any $\hat{m}_\mathrm{max} \gtrsim 10 \mathrm{M}_\odot$ and not close to $\mathrm{m_{up}}$, there is a probability larger than 90\% that the most massive star in the system is larger than such  $\hat{m}_\mathrm{max}$ value. 
Therefore, assuming that Eq.~\ref{eq:last} provides the actual mass of the most massive star in the cluster, as argued in the framework of different interpretations of the IMF,   is an {\it ad hoc} assumption and not a physical fact.

\item $\hat{m}_\mathrm{max}$  defines the mode of the distribution $\Phi_{\cal N}({\cal N}| m_\mathrm{max})$ of the possible $\cal N$ values inferred from the most massive star in the cluster assuming a flat $\Phi_{\cal N}({\cal N})$ distribution. A similar dependence in $\Phi_{\cal N}({\cal N})$ is present when $\cal N$ is inferred from the number of the $N_\mathrm{a}$ most massive stars in the cluster (cf., Paper II). However, the observational evidence is that $\Phi_{\cal N}({\cal N})$ is a power law (if it is related with the ICMF).

\item When the total cluster mass is inferred through the equation $\left<{\cal M}\right> = {\cal N} \times \left< m \right>$ and $\cal N$ is obtained assuming a 
flat $\Phi_{\cal N}({\cal N})$, the observational data become consistent with a $\hat{m}_\mathrm{max} - \left<{\cal M}\right>$ statistical correlation. 
This is indeed the case when $\Phi_{\cal N}({\cal N})$ is not taken into account explicitly in the $\cal N$ (and $\cal M$) estimation (as found in  most of the cluster in the \citealt{WKB10} sample). 

\item The meaningful distribution to be tested against observational data is $\Phi_{m_\mathrm{max},{\cal N}}(m_\mathrm{max}, {\cal N})$ and not $\Phi_{\cal N}({\cal N}| m_\mathrm{max})$ or $\Phi_{m_\mathrm{max}}(m_\mathrm{max}|{\cal N})$.

\item \cite{WKB10} claim that the results of their analysis falsify the hypothesis of a random sampling of the IMF. Based on the two preceding points, we consider that such claim should be revised, both because of the $\cal M$ values it relies on and because of the methodological choice of using $\Phi_{m_\mathrm{max}}(m_\mathrm{max}|{\cal N})$.

\item Different sampling algorithms proposed in the literature are not physical requirements, but convenient mathematical algorithms that try to simplify the implications of such physical law on studies where the IMF is used (as is the case of stellar population in galaxies). Unfortunately, such simplification is not possible.

\item We cannot exclude that a hard physical law linking ${\cal M}$ to $m_\mathrm{max}$ (the {\it actual}  values) does indeed exist; but, if this is the case, it must arise from considerations of the problem including a full-fledged SF analysis, which cannot be shortcut through algebraic IMF manipulations. Whatever the case is, the existence of such an ${\cal M} - m_\mathrm{max}$ physical law is compatible with random sampling of stellar masses and a probabilistic interpretation of the IMF. 

\item If such a physical law exists, it cannot be incorporated to an analytical IMF functional form, but must rather be approached by computing Monte Carlo simulations 
and taking into account only the subset of simulations that verify the assumed ${\cal M} - m_\mathrm{max}$ physical law.
We note that this approach is fully compatible with the optimal sampling {\it definition} provided by \cite{Kroetal11}.

\end{enumerate}

 We conclude that a random sampling IMF is not in contradiction to 
a possible $m_\mathrm{max} - {\cal M}$ physical law. However, such a law cannot be obtained from IMF algebraic manipulation
or included analytically in the IMF functional form. The possible physical information that would be
obtained from the $\cal N$ (or $\cal M$) $-m_\mathrm{max}$ correlation is closely linked with the 
 $\Phi_{\cal M}({\cal M})$ and $\Phi_{\cal N}({\cal N})$ distributions;
hence it depends on the SF process and the assumed definition of stellar cluster. In a second paper of this series we will explore the application of the probabilistic description of the IMF formulated in this study. Particularly, we will describe how to use it to make inferences about quantities that characterize some stellar systems, and how observational constraints work as a priori conditions, affecting the sampling distributions of $\cal M$ and $\cal N$ that we can infer.

\appendix
\section{The intensity function}
\label{subsubsect:IntFunc}

As stated in Sect.~\ref{sec:Ntot-mmax}, $\phi(m)$ cannot provide a value of $m_\mathrm{max}$ that can be used as the {\it actual} maximum stellar mass in a hypothetical cluster. Still, we can calculate 
the probability for the actual value of  $m_\mathrm{max}$ to be close to the mean, the median, the characteristic value, or the mode of $\Phi_{m_\mathrm{max}}({m_\mathrm{max}}|{\cal N})$. In general, we can evaluate the probability that a value known to be larger that $m_\mathrm{b}$ is smaller than  $m_\mathrm{b}+\mathrm{d}m_\mathrm{b}$.  To do that, we need to introduce the intensity function\footnote{We use $\mu(m)$ to follow the notation used by \cite{gumbel}. It must not be confused with the definition of the mean value that is used in other papers.}, $\mu(m_\mathrm{b})$:  

\begin{equation}
\mu(m_\mathrm{b}) \;\mathrm{d} m_\mathrm{b} = \frac{\phi(m_\mathrm{b}) \mathrm{d}m_\mathrm{b}}{1- p(m < m_\mathrm{b})} \geq \phi(m_\mathrm{b})\; \mathrm{d}m_\mathrm{b}.
\end{equation}

The intensity function is not a pdf; it  is independent of ${\cal N}$, as implicit in the idd variable hypothesis: the probability of obtaining a value equal to or larger than 5 throwing one dice is 2/6, independently of previous throws. This must not be confused with the case we studied in the previous paragraphs, which would be equivalent to the probability of obtaining {\it at least} one throw with a result equal to or larger than 5 {\it in} ${\cal N}$ draws.

\begin{figure}
\resizebox{\hsize}{!}{\includegraphics{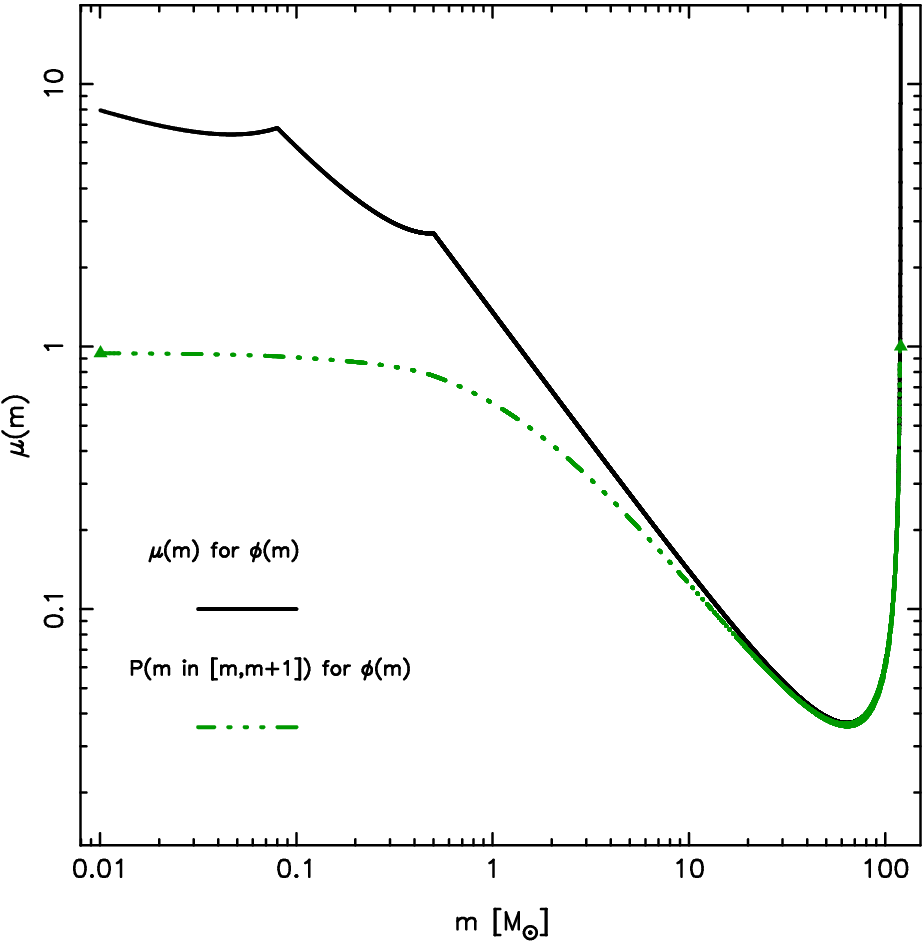}}
\caption[]{Intensity function $\mu(m)$ as a function of $m$ for the IMF. The figure also shows the probability that $m$ will be in the range ($m_\mathrm{b}$, $m_\mathrm{b}+1 \mathrm{M}_\odot$).}
\label{fig:intFunct}
\end{figure}

In Fig.~\ref{fig:intFunct} we plot the intensity function for different values of $m_\mathrm{b}$ for the case of the IMF used in this work. The figure also shows the probability that a star known to have $m\geq m_\mathrm{b}$ will be in the range [$m_\mathrm{b}$, $m_\mathrm{b}+1 \mathrm{M}_\odot$).
The figure shows that  $\mu(m_\mathrm{b})$ has a minimum at a value close to $\mathrm{m_{up}}$, and it goes to infinity at $\mathrm{m_{up}}$. 
The probability of $m$ in the range  [$m_\mathrm{b}$, $m_\mathrm{b}+1 \mathrm{M}_\odot$] decreases with $m_\mathrm{b}$, except for values close to $\mathrm{m_{up}}$. For example, 
there is only a chance lower than 10\% that, given a star in the $m_\mathrm{b} - \mathrm{m_{up}}$ range, this star has a mass $m_\mathrm{b}$ for 
$m_\mathrm{b} \ge 10 \mathrm{M}_\odot$. The situation changes in the extreme case in which  $m_\mathrm{b}$ is close to $\mathrm{m_{up}}$: if we know that 
there is one star with mass $\mathrm{m_{up}}$ or larger, the mass must certainly be $\mathrm{m_{up}}$ (i.e., 
probability equal to 1), since stars with mass larger than $\mathrm{m_{up}}$ do not exist.

This has an interesting implication for the statement that $\hat{m}_\mathrm{max}$ actually provides the mass of the most massive star in the cluster: {\it assuming that  there is one star equal to or more massive than $\hat{m}_\mathrm{max}$ and that $\hat{m}_\mathrm{max} \ge 10 \mathrm{M}_\odot$ and is not close to $\mathrm{m_{up}}$, there is a probability larger than 90\% that the most massive star is more massive than $\hat{m}_\mathrm{max}$!}

\begin{acknowledgements}
MC acknowledges Fernando Selman and David Valls-Gabaud for useful discussions on this subject.  He also acknowledges Roberto Terlevich, Michele Fumagalli, S{\o}ren S. Larsen, and Kevin Covey for discussions on the similarities and differences of $\Phi_{\cal N}({\cal N})$ and $\Phi_{\cal M}({\cal M})$ and their implications in the modeling of clusters and galaxies, which have been very useful for this paper and for future works. Finally, we acknowledge Nate Bastian, Pavel Kroupa, Michele Fumagalli, and John Eldridge for useful comments to the first version of this paper (now split into Papers I and II) and the suggestions of the referee, Peter Anders, which have greatly improved the clarity of the paper. 
This work has been supported by the MICINN (Spain) through the grants AYA2007-64712, AYA2010-15081, AYA2011Ð22614, AYA2010-15196, AYA2011-29754-C03-01, AYA2008-06423-C03-01/ESP,  AYA2010-17631, a Calar Alto Observatory postdoctoral fellowship, and by program UNAM-DGAPA-PAPIIT IA101812, and  CONACYT 152160 Mexico, and co-funded under the Marie Curie Actions of the European Commission (FP7-COFUND)
\end{acknowledgements}

\clearpage

\end{document}